\documentclass[journal]{IEEEtran}
\pdfoutput=1
\usepackage{epstopdf}
\usepackage{multirow}
\usepackage{balance}
\usepackage{algorithm}
\usepackage{algorithmic}
\usepackage{float}
\usepackage{subfigure}
\usepackage{bm}
\usepackage{xcolor}
\usepackage{graphicx}
\usepackage{mathrsfs}
\usepackage{amsmath}
\usepackage{booktabs}
\begin{document}

\newcommand{\revise}[1]{\textcolor{black}{#1}}
\newcommand{\majrev}[1]{\textcolor{black}{#1}}

\title{Synthesizing Brain-Network-Inspired Interconnections for Large-Scale Network-on-Chips}

\author{Mengke~Ge,
        Xiaobing~Ni,
        Qi~Xu,~\IEEEmembership{Member,~IEEE,}
        Song~Chen,~\IEEEmembership{Member,~IEEE,}
        Jinglei~Huang,
        Yi~Kang,
        and~Feng~Wu,~\IEEEmembership{Fellow,~IEEE}
\thanks{This work was partially supported by the National Key R\&D Program of China under grant No. 2019YFB2204800, National Natural Science Foundation of China (NSFC) under grant Nos. 61874102, 61732020, 61931008, and U19A2074, and Strategic Priority Research Program of Chinese Academy of Sciences under grant No. XDB44000000. The authors would like to thank Information Science Laboratory Center of USTC for the hardware \& software services.}
\thanks{Mengke Ge, Xiaobing Ni, and Qi Xu are with the School of Microelectronics, University of Science and Technology of China, Hefei, Anhui, 30332 China. (e-mail: gmk@mail.ustc.edu.cn.)

Song Chen, Yi Kang, and Feng Wu are with the School of Microelectronics, University of Science and Technology of China, and Institute of Artifcial Intelligence, Hefei Comprehensive National Science Center, Hefei, Anhui, 30332 China. (e-mail: songch@ustc.edu.cn.)

Jinglei Huang is with the State Key Laboratory of Air Traffic Management System and Technology, Nanjing, Jiangsu, China.}}

\maketitle
\begin{abstract}
\label{sec:abstract}
Brain network is a large-scale complex network with scale-free, small-world, and modularity properties, which largely supports this high-efficiency massive system.
In this paper, we propose to synthesize brain-network-inspired interconnections for large-scale network-on-chips.
Firstly, we propose a method to generate brain-network-inspired topologies with limited scale-free and power-law small-world properties,  which have a low total link length and extremely low average hop count approximately proportional to the logarithm of the network size.
In addition, given the large-scale applications, considering the modularity of the brain-network-inspired topologies, we present an application mapping method, including task mapping and deterministic deadlock-free routing, to minimize the power consumption and hop count.
Finally, a cycle-accurate simulator $BookSim2$ is used to validate the architecture performance with different synthetic traffic patterns and large-scale test cases, including real-world communication networks for the graph processing application.
Experiments show that, compared with other topologies and methods, the brain-network-inspired NoCs generated by the proposed method present significantly lower average hop count and lower average latency.
Especially in graph processing applications with a power-law and tightly coupled inter-core communication, the brain-network-inspired NoC has up to 70\% lower average hop count and 75\% lower average latency than mesh-based NoCs.
\end{abstract}


\IEEEpeerreviewmaketitle

\section{Introduction}
\label{sec:intro}
Network-on-chip (NoC) \cite{book:Interconnection} is a promising design paradigm for addressing communication bottlenecks in many-core processors.
NoCs replacing point-to-point and shared bus interconnections have been employed to solve complex and large-scale on-chip communication issues because of their scalability, predictability, and modularity \cite{TechPersp,TOBIAS2006A,ASNoC:12,ChenGeneralized,WuAn}.
Recently, it is also widely used in wafer-level integration \cite{wafer} and in-memory computing systems \cite{TrueNorth2014,Neurocube,PIM-DATE,2014Neurogrid}. 
In large-scale NoCs, the interconnection topology has a significant impact on power consumption and performance \cite{book:Interconnection,GLSVLSI}.
TrueNorth \cite{TrueNorth2014,Sawada2016TrueNorth}, a high-efficiency in-memory processor, is composed of 4096 neurosynaptic cores tiled in a mesh.
Xiao et al. \cite{PIM-DATE} present a scalable in-memory computing-based system where hundreds of vaults are interconnected through a mesh-based NoC.
Thousands of cores were integrated in an NoC-based disease diagnosis-on-chip platform for protein folding computation \cite{diseaseDiagnosis}.
Cerebras \cite{wafer}, a wafer-scale engine for deep learning, consists of 400,000 programmable computing cores interconnected by a mesh.
However, these mesh-based interconnections may be unaccommodated for large-scale network architectures,
because NoC with conventional regular topologies generally has an average-hop-count increased, polynomially, with the network size. For example, in a network including $n$ switches connected by mesh, the maximum hop count could be $2n^{1/2}$, which causes an unacceptable global communication latency.
As thousands of cores have been integrated into a single chip for enhanced performance and functionality, network topology becomes a key factor in the success of the large-scale NoCs.

From the perspective of low hop count and high energy efficiency of NoC topology, researchers have carried out in application-specific topology and hierarchical topology respectively.
Some previous studies \cite{WeiZhong,ChenGeneralized,ASNoC:12,Huang2018Lagrangian,Li2018A,Tosun} have proposed topology generation methods of application-specific NoCs.
A series of performance-driven custom topologies were constructed for application requirements.
However, the existing generation methods show their effectiveness and efficiency in applications with hundreds of nodes at most due to high algorithm complexity.
Other studies have proposed the use of hierarchical topologies, such as dubbed Prism-Mesh\cite{Prism-Mesh}, CHMesh\cite{CHMesh}, hierarchical star-mesh \cite{HNoC}, and hierarchical ring \cite{tNoC}, as attractive solutions to the problem of global long-distance data transmission over large-scale networks, due to its lower average hop count compared with conventional regular topologies.
However, for larger network sizes, more high-level switches need to be added to the network, resulting in a large number of switches.

Human brain is a high-efficiency massive parallel computing system combining computing, storage, and communication.
Neurobiological studies have shown that the functional network and structural network of human brain are complex, irregular, and high-efficiency communication networks \cite{Avena2017}.
Although graph theory-based network analysis helps demonstrate the complex organization of human brain networks, how this complexity supports communication processes that are fundamental to the brain's computational capacities remains poorly understood \cite{Avena2017}. It is generally recognized that these two networks have the properties of scale-free \cite{EguScale}, small-world \cite{2017Small}, and modularity \cite{Sporns2016} that to large extent support this high-efficiency system.
Networks in which nodes are connected by a power law of node degree distribution are called $scale$-$free\ networks$  \cite{BAmodel}.
Such networks have extremely low average communication hop count between any pair of nodes, which is logarithmically (from $O(ln$\,$n)$ to $O(lnln$\,$n)$) with network size.
Owing to the heterogeneity of the scale-free networks, when failures of nodes occur randomly, most connections between nodes are conserved, so it is robust against structural breakdown.
$Small$-$world\ networks$ \cite{Watts1998Collective} have a very low average communication hop count between any pair of nodes, which is usually proportional to the logarithm of the network size.
Small-world networks are associated with high global and local efficiency of parallel information processing due to their dense local clustering (modularity).
$Modularity$ means that the nodes in the network naturally are clustered into tightly connected communities with only sparser connections between them \cite{DinhCommunity}.
The modularity of brain networks plays a vital role in promoting stability and flexibility and conserving wiring costs.
Therefore, inspired by efficient brain networks, it is of great interest to build an interconnection topology based on these properties for large-scale NoCs to improve the integration scale and reduce interconnection cost.
Even better than conventional regular topologies, the brain-network-inspired topology has an extremely lower average hop count between any two nodes, which is logarithmically proportional to the network size.
Especially, the brain-network-inspired NoC is more deserving of being explored as a domain-specific solution for graph processing \cite{SOSP-Edge-centric}.
because the large data sets \cite{GraphAnalytics} come from social networks, web pages, bioinformatics, and recommendation systems.
These large networks of datasets are similar to brain networks and follow a pattern of sparsity and power-law distribution \cite{GraphAnalytics}.

Earlier research also showed that better network performance can be obtained by only imitating one of the above properties.
The work \cite{OshidaPacket} generates scale-free NoC based on the Barabasi-Albert (BA) model \cite{BAmodel} without constraints on switch size.
The BA model can only generate a scale-free network of fixed degree distribution exponent ($\approx$ 3).
The growth models including BA model for scale-free networks have been extensively studied in network science.
At every timestep, a new node with several links is added to the initial network which has a small number of nodes.
Huge degree nodes (high-radix switches) impose unacceptable costs due to the power and area of the switch increase superlinearly with its size.
To improve the applicability of scale-free topology for peer-to-peer networks, Hasan Guclu et al. \cite{Guclu2009Limited} and Eyuphan Bulut et al. \cite{BulutConstructing} constructed a limited scale-free network topology with the constraint of maximal node degree, which effectively improved the search efficiency of peer-to-peer networks, respectively.
Unfortunately, in the previous scale-free topology generation methods, the wiring cost \majrev{(link length)} is ignored due to the lack of consideration for the length of links connecting nodes.
\majrev{The link length has a dominant effect on the overall performance and power consumption of NoCs.}
Sourav Das et al. \cite{DasDesign} also designed a 3D small-world NoC architecture based on Watts-Strogatz model according to traffic-driven power-law length distribution, but it is only applied to application-specific NoCs containing tens to hundreds of nodes, and \majrev{its} link length \revise{does not} mathematically obey a power-law distribution.
This classical Watts-Strogatz model \cite{Watts1998Collective} is built from an initial mesh topology and then every link is rewired with a probability to a randomly chosen node.
The rewiring procedure establishes long-range links in the network, which dramatically \majrev{lowers} the average hop count of the topology.
Furthermore, inspired by small-world cortical networks with sparse long-range connections \cite{cortical}, the rewiring of links according to \majrev{the} power-law distribution of link length could further lower the wiring cost associated with communications \cite{Teuscher2007Nature,2005Spatial,DasDesign}.

\revise{Mapping tasks on the topology to achieve low power consumption is another key step in NoC design.
NoC mapping is an NP-hard problem \cite{MAPPING-NPH}.
In \cite{NMAP}, a branch-and-bound algorithm was adopted for application mapping in a regular mesh-based NoC architecture, which minimized the total amount of power consumed in communications.
Tosun et al. \cite{Tosun-map} presented a new integer linear programming (ILP) based application mapping tool for mesh-based NoCs.
Wang et al. \cite{WangWOAGA} proposed a metaheuristic algorithm called WOAGA for large-scale mesh-based NoC mapping to achieve low-energy consumption.
However, \majrev{these methods can be very  time consuming} when the number of tasks reaches several hundred.
For neural network applications, Sawada et al. \cite{Sawada2016TrueNorth} developed a greedy-based algorithm to place tasks one by one on mesh-based TrueNorth based on the Manhattan distance between the current task and the tasks that \majrev{have already} been mapped, which is not suitable for complex task graphs and irregular topologies.
\majrev{A} unified flow combining task scheduling and core mapping \cite{OuHe} is proposed to support regular \majrev{meshes}, irregular \majrev{meshes}, and custom \majrev{NoCs}, using mixed integer linear programming (MILP), then a partition-based speedup technique is proposed to accelerate this model, but it is not suitable for large-scale networks due to \majrev{a} non-polynomial complexity of the MILP problem.
Another partition-based mapping approach \cite{A3MAP} \majrev{was} proposed for large-scale NoCs.
The near-convex region of topologies was selected one by one for each subset of cores obtained by min-cut partitioning on task communication graph.
However, different from \majrev{a} mesh-like topology, the convex regions formed by adjacent cores are not densely connected regions in the brain-network-inspired} topology.
Efficient application mapping for the large-scale complex brain-network-inspired NoC topology is a key problem to be solved urgently in this paper.

The irregularity of the topology induced by the brain-network-inspired design makes packet routing quite complicated.
Zhang et al. \cite{tongji} presented a deterministic-path routing algorithm to tolerate many faults in large-scale NoCs.
Kinsy et al. \cite{BSOR} developed a bandwidth-sensitive oblivious routing approach to produce application-aware \majrev{deadlock-free} routes.
However, these packet routing algorithms did not consider the constraints of both bandwidth and hop count.

\majrev{In this article, we propose to synthesize efficient brain-network-inspired interconnections for large-scale NoCs, considering switch size, link length, and power consumption.
Different from the conference version \cite{GLSVLSI}, this article proposes the topology generation method takes into account the link length constraints to avoid high transmission delays, and further uses a detailed cycle-accurate simulator $BookSim2$ under synthetic traffic patterns and extends $BookSim2$ to support simulation with specific applications, to evaluate the performance of the brain-network-inspired NoC and the effectiveness of the proposed mapping method.}
The key technical contributions of this work are listed as follows.
\begin{itemize}
  \item We propose a method to generate a large-scale brain-network-inspired NoC topology with limited scale-free and small-world power-law properties, which has a low average hop count approximately proportional to the logarithm of network size.
      Then, a modified Louvain algorithm is performed to extract communities of the brain-network-inspired topologies to realize modularity.
  \item Considering the communication requirements of different applications and community structure of the topology, we propose an application mapping method, including task mapping based on the modularity of the brain-network-inspired topologies and Lagrangian relaxation-based deterministic \majrev{deadlock-free} routing scheme, to minimize the communication power consumption and hop count for routing.
  \item Experiments with $BookSim2$ show that the generated brain-network-inspired topologies have significantly lower latency and power consumption under many synthetic traffic patterns compared to other previous topologies. For large-scale applications, the resulting NoC designs show better performance, especially in graph processing, with up to 70\% lower average hop count and 75\% lower average latency than mesh-based NoCs.
\end{itemize}

In the remainder of this paper, Section \ref{sec:motivation} shows \majrev{an} overview. Section \ref{sec:topo-gene}$\sim$\ref{sec:appmap} present the topology generation and application mapping of the brain-network-inspired interconnections for large-scale NoCs, respectively. Section \ref{sec:experiment} lists experiments, followed by \majrev{a} conclusion in Section \ref{sec:conclusion}.

\section{Overview}
\label{sec:motivation}

In this work, the NoC architectures are assumed to support packet-switched communications with source routing and wormhole flow control, \revise{and we choose a switch architecture similar to the input-queued switch architecture \cite{Jiang2013A} with a four-stage pipeline for packet header flits.}
Each core is connected to a switch, and we focus on the network topology between switches, where each node represents a switch.
\begin{figure}[htbp]
\small \centering
  \includegraphics[width=7.5cm]{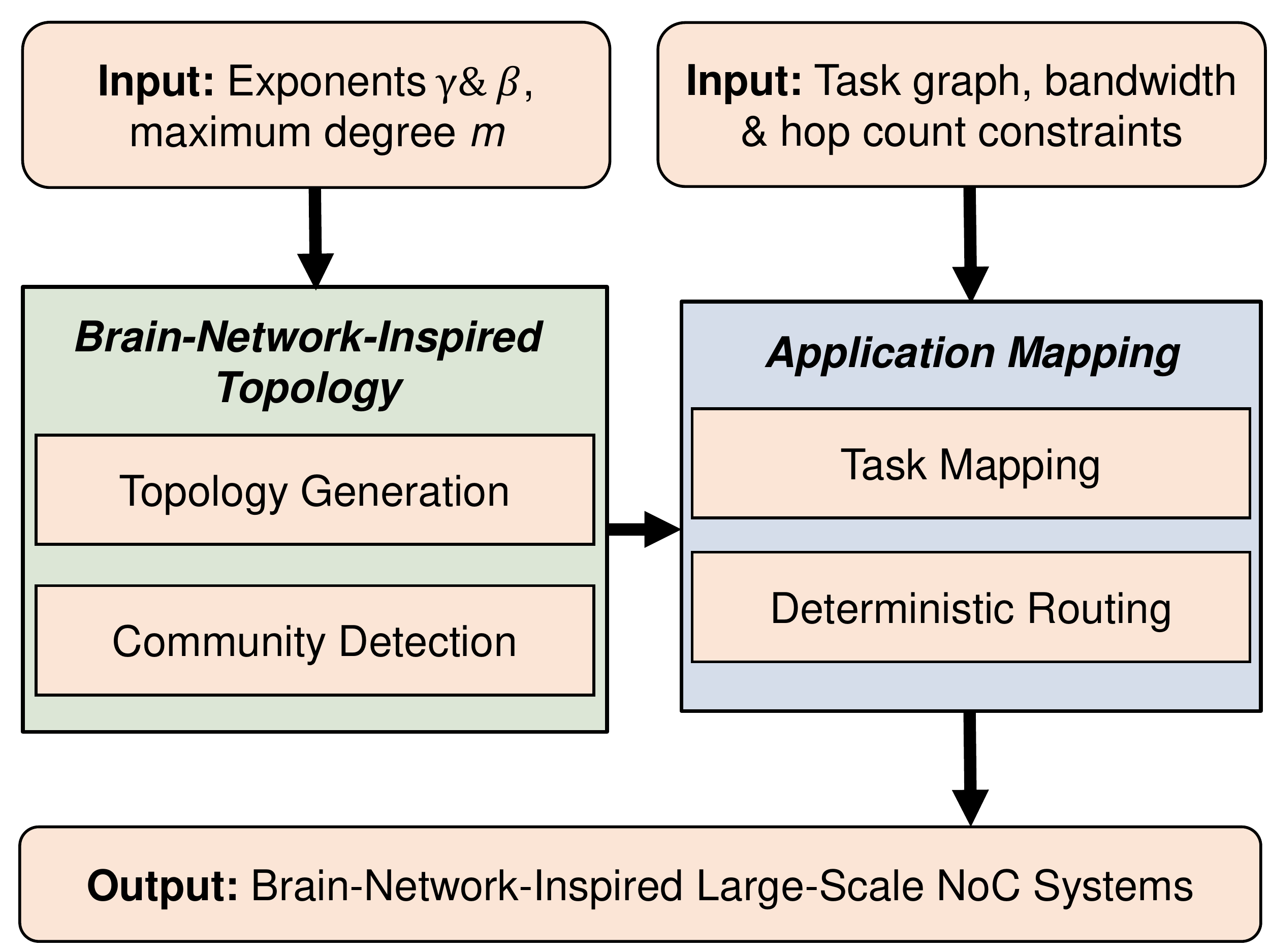}\\
  \caption{\revise{Design flow.}}
  \label{fig:flow}
\end{figure}

In our design, we first propose to generate the brain-network-inspired interconnection topology with a low average hop count approximately proportional to the logarithm of network size for large-scale NoCs, then propose to address the large-scale application mapping problem for this brain-network-inspired NoCs.
Figure \ref{fig:flow} shows the design flow.
In \majrev{the} first stage, given power-law distributions, a deterministic growth algorithm is proposed to construct the brain-network-inspired topology with limited scale-free and small-world power-law properties.
\majrev{To obtain a proper brain-network-inspired topology, we evaluate the communication-independent basic power consumption and communication cost of all topologies generated with different power-law distributions of node degree and link length and make a tradeoff.}
Then, a modified Louvain algorithm is performed to extract communities of the brain-network-inspired topologies to realize modularity.
In task mapping of the second stage, given the specific communication requirements of the application, we exploit the modularity of the brain-network-inspired topologies to solve the difficulty in large-scale task mapping.
\majrev{We first propose a $k$-way partition and simulated annealing based heuristic method to assign the tasks with heavy traffics to the same community in the topology, and then present a detailed task placement method to place each task to a specific core one by one.}
In \majrev{the} deterministic routing of the second stage, given the constraints of hop count and bandwidth, we develop an application-specific Lagrangian relaxation-based \majrev{deadlock-free} routing scheme on the brain-network-inspired NoC, which statically determines routes for all flows.
A detailed cycle-accurate simulator, $BookSim2$ \cite{Jiang2013A}, is applied to measure latency, power consumption, and hop count metrics that aid a quantitative, well-informed comparison between multiple NoC systems.
We extend $BookSim2$ to simulate communication architectures with specific applications according to the solution of application mapping. For convenience, key notations used in this paper are listed in Table \ref{tab:notation}.

\begin{table*}[htbp]
\centering
  \caption{Key notations used in this paper.}
  \label{tab:notation}
  \renewcommand\tabcolsep{5.0pt}
  \begin{tabular}{|c|p{14cm}|}
    \hline
    $\gamma$ & Exponent of power-law distribution of node degree, and \majrev{$\gamma>0$}\\
    \hline
    $\beta$ & Exponent of power-law distribution of link length, and \majrev{$\beta>0$}\\
    \hline
    $i$ & Node degree, and \majrev{$k\leq i \leq m_a$}\\
    \hline
     $m$ &  \majrev{User-defined maximal switch size, and $m>2k$}\\
    \hline
    $m_a$ &  \majrev{Actual maximal node degree, and $2k<m_a \leq m$}\\
    \hline
    $k$ & Initial node degree, and \majrev{$k\geq 2$}\\
    \hline
    \majrev{$l_a$} & \majrev{User-defined maximal link length}\\
    \hline
    $l$ & Link length, \majrev{and $0<l\leq l_a\leq 2(t-1)$}\\
    \hline
    $n$ & Network size, \majrev{and $t*t=n$}\\
    \hline
    $G_s(S,E_s)$ & \revise{Brain-network-inspired} topology in which each vertex $s_u\in S$ represents a core/switch \majrev{($|S|=n$)}, each edge $(s_u,s_v)\in E_s$ represents a link from core/switch $s_u$ to $s_v$, and the length of link $(s_u,s_v)$ is given by $\omega_{uv}$.\\
    \hline
    $G_t(T,E_t)$ & Task graph in which each vertex $t_u\in T$ represents a task and each edge $(t_u,t_v)\in E_t$ with the weight $cr_{uv}$ represents the communication requirement from task $t_u$ to $t_v$.\\
    \hline
    $G_c(C,E_c)$ & Core communication graph in which each vertex $c_u\in C$ represents a core/switch and each edge $(c_u,c_v)\in E_c$ with the weight $fl_{uv}$ represents the communication requirement from core/switch $c_u$ to $c_v$.\\
    \hline
\end{tabular}
\end{table*}

\section{Generating A Brain-Network-Inspired Topologies for Large-Scale NoCs}
\label{sec:topo-gene}
\subsection{Topology Generation}
\label{sec:topo-sel}
\revise{\noindent\textbf{Problem statement:} Given network size $n$, maximal switch size constraint $m$,  \majrev{and maximal link length constraint $l_a$}, we attempt to generate a \revise{brain-network-inspired} topology $G_s(S,E_s)$ with the properties of limited scale-free and power-law small-world for large-scale NoCs, which has low power consumption and low average hop count between any two nodes approximately proportional to the logarithm of the network size.}

In this work, we propose a deterministic growth algorithm by introducing switch size \majrev{and link length} constraints, \majrev{which is different from the semi-deterministic growth algorithm} (SDA) \cite{BulutConstructing}, to generate a brain-network-inspired NoC topology with limited scale-free and power-law small-world properties, which has a \majrev{low total link length}, low power consumption, and low average hop count between any two nodes approximately proportional to the logarithm of the network size.
\majrev{Given different power-law distributions, the deterministic growth algorithm can generate different topologies. In order to select a proper topology, we evaluate the communication-independent basic power consumption and communication cost of the topology and make a tradeoff.}
Algorithm \ref{alg:toposelect} shows the key steps of topology generation.

Limited scale-free means that \majrev{the maximum node degree (switch size) of the network cannot be greater than the user-defined maximal switch size $m$ and the node degree follows a power-law distribution $P(i)\propto i^{-\gamma}$ only if the node degree $i$ is less than $m$}.
In addition, the authors \cite{Teuscher2007Nature,2005Spatial,DasDesign} have shown that networks with a power-law distribution rather than a uniform distribution of link length \majrev{can not only result in the small-world property, but also} physical realizability and low total link length.
Furthermore, while seeking a power-law distribution of node degree in the process of network growth, it is another pursuit of topology generation to realize the power-law distribution of link length, $P(l)\propto l^{-\beta}$, \majrev{where $l$ is link length.
To avoid generating a very long link, which has extremely high transmission delay, there is a user-defined maximal link length $l_a$, and the power-law distribution of link length exists only if $l<l_a$.}
Given $\gamma$ and $\beta$, our method generates a network topology deterministically by growth, so we call it deterministic growth algorithm.
By varying $\gamma$ and $\beta$, more topologies can be generated.

\majrev{Power consumption of NoCs, including basic network power consumption and communication cost, is an important index in network topology design.
Basic power consumption, including static power consumption and clocking power consumption, is communication-independent.
In synchronous design, given the network topology, there is a certain basic power consumption.}

\majrev{To select a proper topology for large-scale NoCs, that is, select a set of $\gamma$ and $\beta$ from many combinations, we evaluate the basic power consumption and communication cost of all topologies.}
Due to the emergence of large-size switches and long-range links that insert repeaters to reduce interconnect transmission delays, the \majrev{basic} power consumption of the brain-network-inspired topology would be greater than that of the conventional mesh, but due to the lower average hop count between nodes, the \majrev{communication} cost of this topology would be lower.
Selecting smaller $\gamma$ and $\beta$ may reduce basic power due to fewer high-degree nodes and long-range links, but it would cause \majrev{a} bigger hop count and larger routing path length during communication, which increase \majrev{communication} cost.
We vary $\gamma$ and $\beta$ by \majrev{the} step size of 0.1, and call the deterministic growth algorithm to generate a topology for each set of $\gamma$ and $\beta$.
Finally, we select a proper topology by \majrev{a tradeoff between} the \majrev{basic} power consumption and \majrev{communication cost} of topologies.

\begin{algorithm*}
\footnotesize
\caption{\textit{Topology\ Generation}}
\label{alg:toposelect}
\begin{algorithmic}[1]
\REQUIRE{Network size $n$, maximal switch size constraint $m$, \majrev{and maximal link length constraint $l_a$}}
\ENSURE{Construct a proper \revise{brain-network-inspired} topology $G_s(S,E_s)$ with limited scale-free and small-world power-law properties}
\STATE Create and initialize a set of topologies $ST$;
  \FOR{$\gamma$ $\leftarrow$ ${\gamma}_{min}$ to ${\gamma}_{max}$}
    \FOR{$\beta$ $\leftarrow$ ${\beta}_{min}$ to ${\beta}_{max}$}
        \STATE Call \textbf{\textit{Deterministic\_Growth\_Algorithm($n$, $m$, \majrev{$l_a$}, $\gamma$, $\beta$)}} and obtain a resulting topology $G_{\gamma \beta}$;
        \STATE Calculate $C_{\gamma\beta}$ and $P_{\gamma\beta}$ of $G_{\gamma \beta}$;
        \STATE Add $G_{\gamma \beta}$ to $ST$;
    \ENDFOR
  \ENDFOR
   \STATE Select a proper topology $G_s(S,E_s)$ with maximal $OBJ$  from $ST$ according to Equation \ref{eq:obj-topo};
\end{algorithmic}
\end{algorithm*}

\majrev{Basic} power consumption is the sum of power consumed by all switches and physical links with repeaters.
The power model \cite{ICS} is used to estimate the basic (static and clocking) power consumption based on the degree of nodes (switch size) and the length of links.
A detailed evaluation of the communication cost for each topology at the actual traffic can be very time-consuming.
As a rough estimate, we assume that communication cost comes from communication between every two-node pairs.
\majrev{We make trade-offs between communication-independent basic power consumption and communication cost to get a proper topology with a maximum of $OBJ$}, which can be expressed as
\begin{equation}
\label{eq:obj-topo}
OBJ=\alpha\cdot\frac{C_{min}}{C_{\gamma\beta}}+(1-\alpha)\cdot\frac{P_{min}}{P_{\gamma\beta}}
\end{equation}
where $P_{\gamma\beta}$ and $C_{\gamma\beta}$ represent respectively \majrev{basic} power consumption and \majrev{communication cost} of the topology with a set of $\gamma$ and $\beta$.
Since a smaller hop count means less \majrev{communication cost} over switches, and shorter routing path length means less \majrev{communication cost} over physical links, we roughly estimate $C_{\gamma\beta}$ to be $\sum_u{\sum_v{L_{u,v}\cdot H_{u,v}}}$, where $H_{u,v}$ and $L_{u,v}$ are respectively \majrev{the} hop count and path length (the sum of the Manhattan distance of links) of the minimum hop path between every two-node pair $(s_u,s_v)$, $\forall s_u, s_v\in S$.
The parameter $\alpha$ is normally set to 0.5, when the \majrev{basic} power consumption of the generated topology is too large to exceed a threshold, such as \majrev{1.3} times of that of mesh, we will reduce $\alpha$ to adjust the relative weight until the \majrev{basic} power consumption of the selected topology is below this threshold.
\majrev{Since the basic power consumption and the rough communication cost have different units of measurement}, for more accurate assessment, $C_{min}$ and $P_{min}$ are respectively minimum communication cost and basic power consumption over all possible cases of topologies for \majrev{the} normalized process.

\subsubsection{Deterministic Growth Algorithm}
\label{sec:deterministic}
Similar to the growth algorithms of complex network models, our deterministic growth algorithm starts with \majrev{an} initial simple topology and adds new nodes in sequence from near to far from the partial network until all nodes have been added. Algorithm \ref{alg:topogen} shows the key steps of the deterministic growth algorithm.
Indeed, in \majrev{the} growth process, the establishment of every link deterministically makes the topology tend to be scale-free and small-world.
During the growth, we first keep the power-law distribution of node degrees of topologies to achieve the scale-free property as much as possible.
In addition, \majrev{links are established under the constraint of link length and tend to be power-law distributions of length}.

\begin{figure}[htbp]
\small \centering
  \includegraphics[width=6cm]{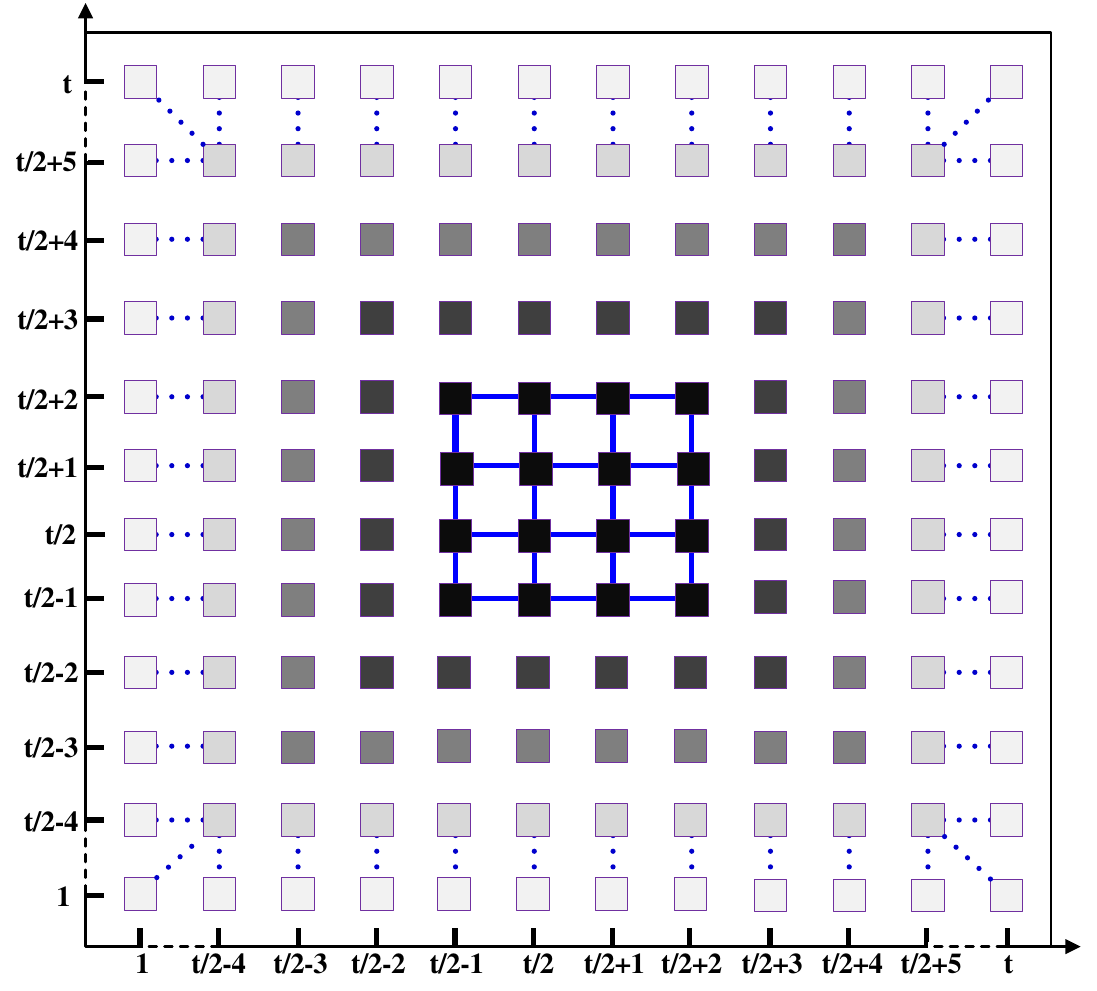}\\
  \caption{On-chip floorplan with an initial mesh topology.}
  \label{fig:floorplan}
\end{figure}

\revise{In general, the initial small topology can be any small connected topology, and the form of the initial small topology has little influence on the power-law distributions of the final large-scale topologies. In this design, we employ a small mesh as the initial topology as shown in Figure \ref{fig:floorplan}.}

\noindent\textbf{Initial Setup:} The NoC design consists of a 2-D grid of $t\times t(=n)$ cores. We build a two-dimensional coordinate system on the plane of the chip, each core has a unique coordinate.
The Manhattan distance of all possible links is within the range $[1, 2t-2]$. 
The degree of all nodes is within $[k, m]$. The interconnection starts with a small-scale topology consists of black cores and blue edges at the center of the chip plane.

\begin{algorithm*}
\footnotesize
\caption{\textit{Deterministic\_Growth\_Algorithm($n$, $m$, \majrev{$l_a$}, $\gamma$, $\beta$)}}
\label{alg:topogen}
\begin{algorithmic}[2]
    \STATE \revise{$m_a=m$;}
    \WHILE{\revise{the inequation \ref{eq:ineq} does not hold}}
                \STATE{\revise{$m_a--$;}}
    \ENDWHILE
    \FOR{degree $i\leftarrow k$ to \revise{$m_a$}}
           \STATE Calculate $f_i$, $freq[i]$, $n[i]$, and $score[i]$;
    \ENDFOR

  \STATE Construct an initial simple small topology in the center of the chip plane;
    \FOR{\revise{each new node $NN$ that has not been yet connected to the topology in sequence from near to far from the partial topology}}
       \FOR{\revise{$lc\leftarrow 1$ to $k$ for $NN$}}
        \STATE \revise{$D_{min} = \infty$, $i_{exp}= 0$};
        \STATE //**\textit{To determine the desired degree of existing nodes for scale-free property.}**//
            \FOR{degree $i\leftarrow k$ to \revise{$m_a$}}
            \STATE \revise{Calculate the difference of absolute deviation $D(i)$ according to Equation \ref{eq:sum-difference}};
            \IF{$\revise{D_{min} > D(i)}$ \&\& $n[i]!=0$}
                \STATE{\revise{$D_{min} = D(i)$, $i_{exp} = i$}};
            \ENDIF
        \ENDFOR
        \STATE \revise{$DI_{max}= 0$, $N_{dir}= 0$};
        \FOR{each node \revise{$EXN$} of the partial topology}
            \IF{the degree of \revise{$EXN$} equals $i_{exp}$ \&\& no existing link between the these two nodes}
                \STATE //**\textit{Establish a new link to satisfy link distribution of power-law small-world property.}**//
                \STATE Calculate the Manhattan distance $l$ \revise{between $NN$ and $EXN$};
                \IF{\majrev{$l>l_a$}}
                    \STATE{\majrev{Continue;}}
                \ENDIF
                \STATE \revise{Calculate the deviation $DI(l) = P_{expected}(l) - P_{actual}(l)$};
                \IF{$DI_{max} < DI(l)$}
                    \STATE{\revise{$DI_{max} = DI(l)$, $\revise{N_{dir}= EXN}$}};
                \ENDIF
            \ENDIF
        \ENDFOR
        \STATE Update $freq[i]$, $n[i]$, and $score[i]$;
        \STATE \revise{Connect $NN$ to $N_{dir}$ by a new link};
        \ENDFOR
    \ENDFOR
\end{algorithmic}
\end{algorithm*}

\noindent\textbf{Growth Strategy:}
During the growth process, each remaining new node is added orderly by connecting $k$ different nodes of the partial topology, and these $k$ nodes of the partial topology are determined one by one for \majrev{the} new node, while ensuring the expected power-law distributions for node degree and link length.
\majrev{Except for the nodes of the initial topology, the remaining new nodes are connected to the network through $k$ links, and the nodes of the initial topology only account for a small proportion of all nodes.}
The initial degree of new nodes is $k$ and each newly established bidirectional link will increase the total degree by 2, so the average degree of all nodes \majrev{in the final network topology is almost $2k$}.
In Line 6-14, we first determine the expected degree of the node of the partial topology to connect the new node, according to the deviation of the power-law distribution of degrees between before and after connecting the new node to a node with each possible degree, and these nodes with the expected degree are considered as candidates.
When the new node connects a node with degree $u$ of the partial topology, it can make the node degree distribution \majrev{closer} to the power law, that is, there is a smaller deviation from the ideal power-law distribution, then $u$ is called the expected degree.

According to \cite{BulutConstructing}, \majrev{the average degree of all nodes is assumed to be $2k$ in the topology, and $m_a$ is the actual maximal node degree, and node degree has a power-law degree distribution $P(i)\propto i^{-\gamma}$, $k\leq i<m_a \leq m$, so $m_a>2k$ and the frequency $f_i$ of nodes with degree-$i$ derived} can be written as
\begin{equation}
\label{eq:f_i}
  f_i=\frac{m_a-2k}{i^{\gamma}\sum_{j=k}^{m_a-1} \frac{m_a-j}{j^{\gamma}}} \quad for\ k\leq i < m_a
\end{equation}
\begin{equation}
\label{eq:f_m}
  f_{m_a}=1-\sum\nolimits_{j=k}^{m_a-1}{f_i}
\end{equation}
So given $m$, $k$, and $\gamma$, we can obtain the expected degree distribution.
In addition, in order for all frequencies $f_i$, $k\leq i\leq m_a$, to be positive, the following inequality must be satisfied \cite{BulutConstructing}:
\begin{equation}
\label{eq:ineq}
2k\cdot \sum_{j=k}^{m_a-1}{i^{-\gamma}} \geq \sum_{j=k}^{m_a-1}{i^{-\gamma+1}}
\end{equation}
\majrev{The initial value of $m_a$ equals $m$, when this inequality is not met, we reduce $m_a$ until it is met. Hence, $m_a\leq m$, and the maximum node degree does not exceed the user-defined constraint of switch size.}

The next step is to determine what degree of nodes to connect to minimize the deviation of the resulting degree distribution from the expected power-law degree distribution. The deviation of degree-$i$ nodes' expected count and current count in the topology is shown as
\begin{equation}
\label{eq:difference}
 d(i)=n[i]-score[i]-freq[i]
\end{equation}
where $n[i]$ and $score[i]$ denote the current node count and expected count of degree-$i$ at a given node count, respectively, and $freq[i]=f_i/k$ is the expected increment in degree-$i$ node count with only one edge addition to a new node. \revise{When connecting to the node with degree-$i$, the deviation of degree-$i$ and degree-($i+1$) varies.}
Consequently, the total difference of absolute deviation of all degrees, between before and after connecting to the degree-$i$ node, can be computed as
\begin{equation}
\label{eq:sum-difference}
 D(i)=(|d(i+1)+1|+|d(i)-1|)-(|d(i+1)|+|d(i)|)
\end{equation}
\revise{The total difference may be negative, and the smaller the value, the smaller the overall deviation. Therefore, \majrev{if the new node connects the degree-$i_{exp}$ node with the smallest total difference, then $i_{exp}$  is the expected degree, and connecting the node with degree-$i_{exp}$ can adhere the power-law distribution of node degree} as much as possible.
Note that when there is only one node with expected degree in the partial topology, we connect the new node to this node directly.}
But it is more likely to have a batch of candidates that satisfy this term.

\revise{Subsequently, in Line 16-25, only one node of candidates is selected to connect the new node according to the deviation between the expected and \majrev{the} actual distribution of link length.}
To further satisfy a power-law distribution of link length,
we select one of these candidates whose length to this new node has the maximum deviation between the expected count and current actual count.

The expected probability of link length can be \majrev{expressed} as
\begin{equation}
\label{eq:sw}
\majrev{P(l)=
 \left\{
    \begin{array}{ll}
        \frac{l^{-\beta}}{\sum_{L=1}^{2(t-1)}{L^{-\beta}}}   &, \ 1\leq l \leq l_a-1\\
        \\
        1-\sum_{L=1}^{l_a-1}{P(L)}   &, \ l = l_a\\
    \end{array} \right.\\}
\end{equation}
where $l$ is computed by the Manhattan distance from the new node to the target node in the coordinate depends on the mainstream metal Manhattan routing (i.e., only allow horizontal and vertical connections).
\majrev{To constrain the length of links ($l \leq l_a$) and preserve the proportion of long links, we set the expected probability of link length $l_a$ to the probability of links whose length is not less than $l_a$ when there is no length constraint, $P(l_a)=1-\sum_{L=1}^{l_a-1}{P(L)}$.}
If there are still multiple nodes satisfying both degree and link length conditions, the existing node with the minimum coordinate is finally selected.

\revise{Finally, the new node is greedily and deterministically connected to a node of the partial topology, which makes the network topology most adherent to the limited scale-free and small-world power-law properties.}
\majrev{Note that the $k$ nodes connected to each new node must be different, and when connecting an existing node to the current new node}, we make sure that there are no existing links between the current new node and the existing node.
Repeat the above steps until all nodes are added to the network, and finally we obtain a brain-network-inspired topology.

\begin{figure*}[htbp]
\centering
\subfigure[]{
	\centering
	\includegraphics[height=3.7cm]{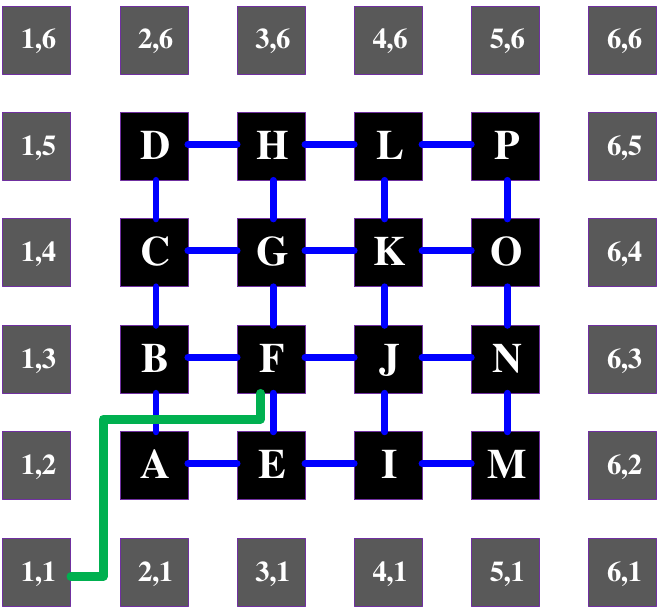}
    \label{fig:topo-a}
}
\hspace{5ex}
\subfigure[]{
	\centering
	\includegraphics[height=3.7cm]{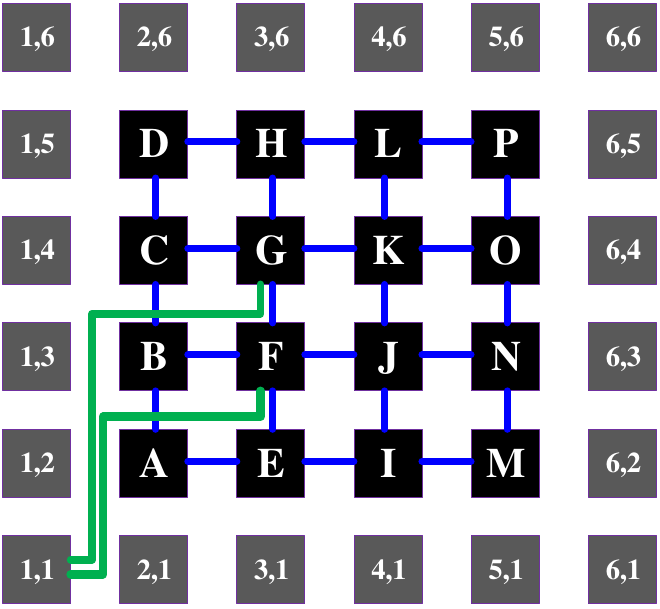}
	\label{fig:topo-b}
}\\
\subfigure[]{
	\centering
	\includegraphics[height=3.7cm]{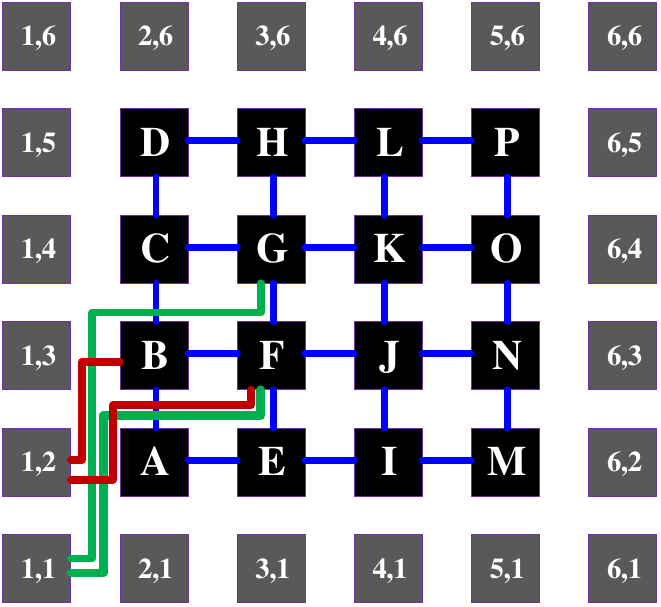}
	\label{fig:topo-c}
}
\hspace{5ex}
\subfigure[]{
	\centering
	\includegraphics[height=3.7cm]{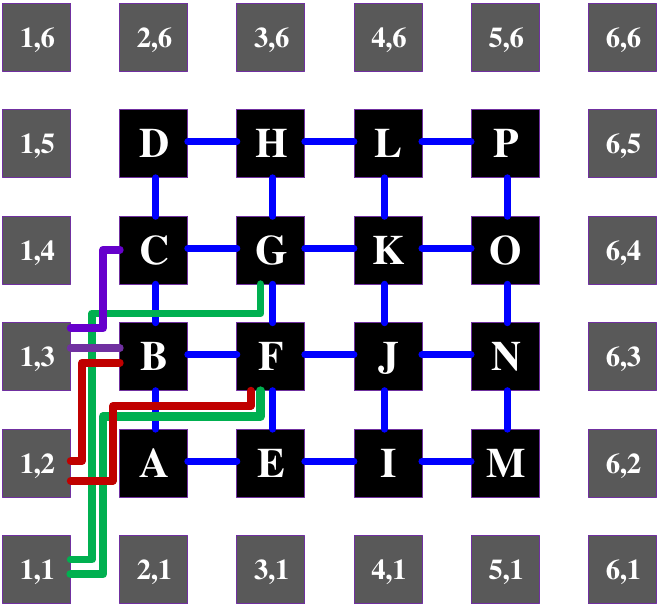}
	\label{fig:topo-d}
}
	\caption{\revise{An example of deterministic growth algorithm.}}
\label{fig:topoexample}
\end{figure*}

\revise{\noindent\textbf{Example:}} \revise{Figure \ref{fig:topoexample} illustrates the process of this method. Nodes $A\sim P$ and the blue edges form an initial topology, and the remaining new nodes [1, 1] $\sim$ [6, 6] are added to the network one by one in order. Each new node connects 2 nodes of the partial topology initially.
Node [1, 1] is added first, at this point, $D(2)$, $D(3)$, and $D(4)$ are respectively 2.0, 0, and -2.0, so $F$, $G$, $J$, and $K$ with degree 4 can be selected as candidates.
The Manhattan distance from node [1, 1] to $F$, $G$, $J$, and $K$ are 4, 5, 5, and 6, respectively.
The deviation between the expected count and current actual count of link with length 4 is the largest, so node $[1, 1]$ is connected to $F$ as shown in Figure \ref{fig:topo-a}.
Then the second node of the partial topology will be selected for node [1, 1], and $D(2)$, $D(3)$, and $D(4)$ are respectively 2.0, 0, and -0.6, so node $G$, $J$, and $K$ with degree 4 can be selected as candidates.
The deviation between the expected count and current actual count of link with length 5 is the largest, so node [1, 1] is connected to $G$ as shown in Figure \ref{fig:topo-b}.
In the same way, we establish two links for node [1, 2] in Figure \ref{fig:topo-c} and node [1, 3] in Figure \ref{fig:topo-d}, respectively.
Repeating the same process until the topology is generated for all remaining nodes.}

\subsection{Community Detection}
\label{sec:community}
Communities are sets of highly interconnected nodes in networks, and the nodes within different communities are only sparsely connected.
\majrev{A fast greedy Louvain algorithm \cite{blondel2008fast} is the most popular method to detect the modularity of complex networks and does not change the topology structure.
In this stage, to avoid oversized communities, which will degrade the usefulness of partition, we modify Louvain algorithm \cite{blondel2008fast} to detect communities of the brain-network-inspired topology by introducing a community size constraint.}
$G_s(S,E_s)$ is the input, but the weight of each edge is given by $1/\omega_{uv}$.

\majrev{Louvain algorithm is an iterative method in which modularity, which measures the density of links inside communities as compared to links between communities and has been used to compare the quality of the partitions \cite{blondel2008fast}, is the optimization objective.
Each community initially includes a node of the topology.
In each iteration, each community is assumed to be in turn merged into its neighboring communities, and then the community is merged into the neighboring community with the largest modularity gain.
Ultimately, modularity does not increase, and each community detected contains tightly connected nodes in the topology.
To avoid oversized communities, in iterations, if a community is merged with one of its neighboring community and the overall size is greater than $TD$, this case would be ruled out.
However, limiting the community size can result in a decrease in modularity and partition quality.
Therefore, when setting TD for topologies of different sizes, it is necessary to ensure that the modularity of community detection cannot be significantly reduced.}

\begin{figure}[htbp]
\small \centering
  \includegraphics[width=6.5cm]{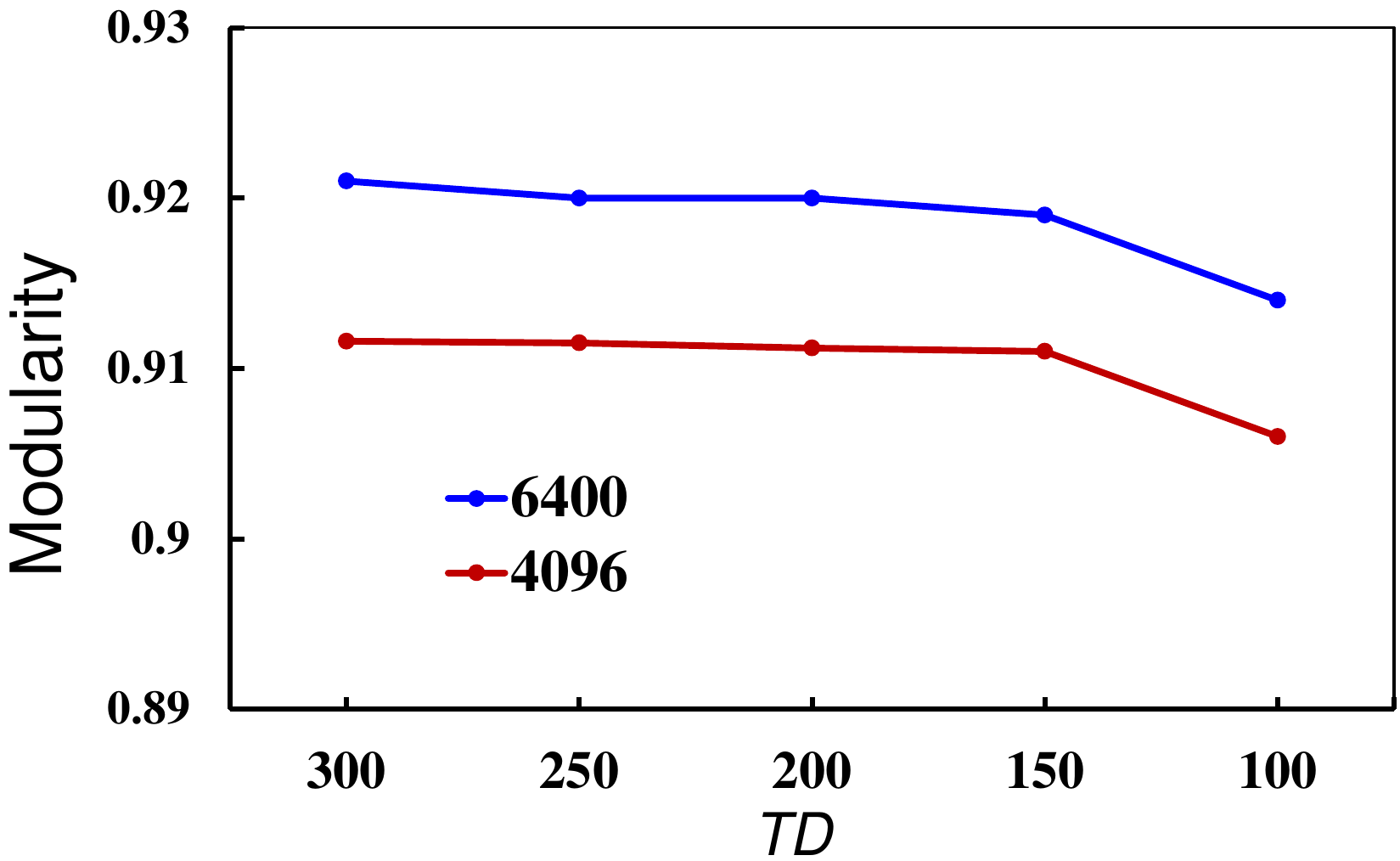}
  \caption{The modularity w.r.t. $TD$.}
  \label{fig:td}
\end{figure}

\majrev{Examples of community detection for brain-network-inspired topologies with 4096 nodes and 6400 nodes are shown in Figure \ref{fig:td}.
The modularity of the partition is a scalar value between -1 and 1.
For these two topologies, after the Louvain algorithm is executed, the topology will be divided into several communities and even some communities have 300 to 500 nodes.
When $TD$ is as low as 150, the modularity does not decrease significantly, otherwise it will degrade the partition quality.}

Finally, all vertices $s_u\in S$ are divided into several communities by the modified Louvain algorithm, and we obtain the set of communities $CM=\{cm_x|\sum_{x}|cm_x|=|S|\bigwedge|cm_x|\leq TD,x=1,2,...,N_M\}$ and the number of communities $N_M$.

Within the framework of network science, high-degree nodes that are positioned to make strong contributions to global brain network function are generally referred to as hubs \cite{hubs2007,hubs2013,natureHUB}. We regard that these hubs act as key feature nodes and reflect the location of all nodes in their communities, which will play an important role in the subsequent application mapping process. We adopt the basis of hub classification \cite{hubs2007,hubs2013,natureHUB} based on the network's community structure in brain network research.
To classify hubs for each community, each node's participation index $P$ \cite{hubs2007,natureHUB} which expresses its distribution of inter- versus intra-community connections is calculated.
$P$ of node \revise{$s_u$} is defined as
\begin{equation}
\label{eq:participation}
\revise{P_{s_u}=1-\sum\nolimits_{s=1}^{N_M}{(\frac{\kappa_{ux}}{i_u})^2}}
\end{equation}
where $N_M$ is the number of identified communities, \revise{$i_u$} is the degree of node \revise{$s_u$}, and $\kappa_{ux}$ is the number of edges from node \revise{$s_u$} to nodes within community \revise{$cm_x$}.
\revise{Smaller $P$ means that the more edges of the node are connected to nodes in the same community.
In the study of brain networks, when participation coefficient  $P < 0.3$, high-degree nodes (at least one standard deviation above the network mean) are defined as provincial hubs which have the vast majority of links within their module \cite{hubs2007,natureHUB}.}
Therefore, in each community, we classify high-degree nodes according to this criterion as hubs.
If there are no nodes in the community that meet the above conditions, we select the three nodes with the highest degree as hubs in this community.
\revise{We define the set of hubs in community $cm_x$ as $HB_x=\{hb_1,hb_2,...\}$, $x=1,2,...,N_M$.}

\subsection{Complexity Analysis}
In summary, topology generation has running time $O(|S|^2+|S|+|E|)$ $(= O(|S|^2))$ , where $|S|$ is the number of nodes, and $|E|$ denotes the total number of edges in the brain-network-inspired topology. Meanwhile, the complexity of community detection \cite{blondel2008fast} and hub classification \cite{hubs2007} for complex topologies has been proved within $O(|S|+|E|)$.

\section{Application Mapping}
\label{sec:appmap}
\subsection{Task Mapping}
\label{sec:taskmap}
\revise{\noindent\textbf{Problem statement:} Given a task graph $G_t(T,E_t)$ and a large-scale \revise{brain-network-inspired} topology $G_s(S,E_s)$, we attempt to map all $|T|$ tasks to the cores of this topology and ensure low communication power and low communication hop count.}

\begin{figure}[htbp]
\small \centering
  \includegraphics[width=5cm]{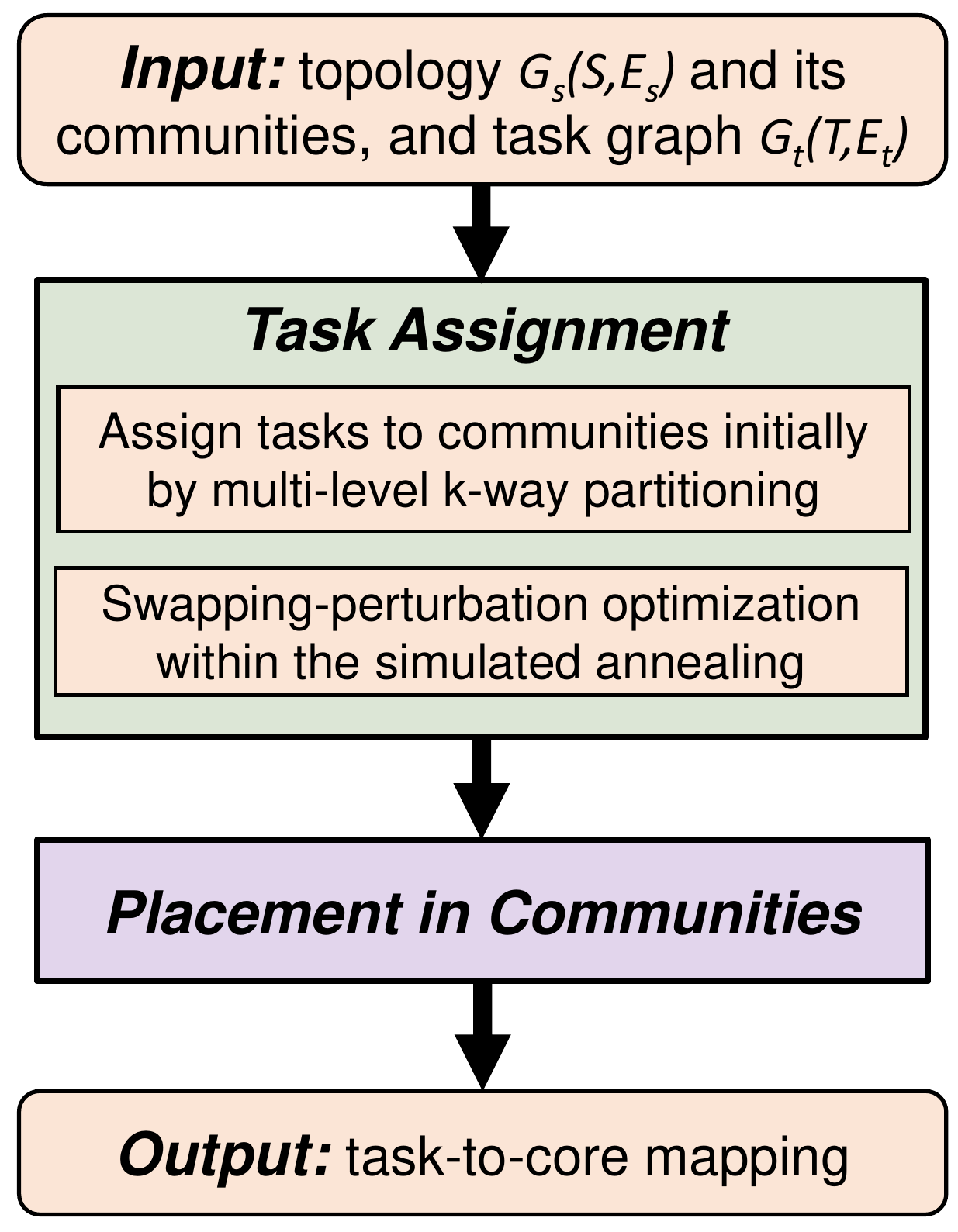}\\
  \caption{\revise{Flow of task mapping.}}
  \label{fig:map}
\end{figure}

Drawing on the modularity of the brain complex network, we exploit community structures to improve the solution quality in large-scale task mapping.
The detailed flow is shown in Figure \ref{fig:map}.
The brain-network-inspired topology has been decomposed into several communities of densely interconnected nodes in the community detection stage.
\majrev{We first propose a $k$-way partition and simulated annealing based heuristic method to assign the tasks with heavy traffics to the same community in the topology.
Then, according to the tasks-to-community assignment, we present a detailed task placement method to place each task to a specific core one by one.
The hubs of the community to which unmapped tasks are assigned are introduced to evaluate the communication cost between the current task and the unmapped tasks, to further improve the quality of the mapping.}

\subsubsection{\majrev{Task Assignment}}
Firstly, according to the task graph $G_t(T,E_t)$, we perform the multilevel $k$-way partitioning algorithm to initially assign tasks into $N_M$ communities.
If the given number of tasks \majrev{is} less than $n$, we add some spare tasks without traffic into the graph.
To ensure that the number of tasks matches the size of each community, the community detection results including the number and size of each community are used as the inputs of partitioning.
We define $N_M$ subsets of tasks as $sub_x$, and ensure $|sub_x|=|cm_x|,x=1,2,...,N_M$.

Furthermore, a perturbation method within the simulated annealing is to randomly choose two tasks in two different communities, swap them, and create a new partitioning for exploring a \revise{superior task-to-community assignment} solution.
\revise{According to \cite{DasDesign}, $o = (r \cdot h+d)\cdot cr$ is used to roughly evaluate the communication cost of each flow, where $r$ denotes the number of switch stages, $h$ is the hop count of each flow, and $d$ is the Manhattan distance between \majrev{the} source node and sink node of each flow.}
Low $o$ means low communication hop count and power consumption.
Since the specific location of tasks is not determined, \revise{to obtain a superior partitioning solution, during the partitioning process the hubs in each unmapped community are regarded to integrate the external communication requirements of tasks within this community.
We define $O$ as the communication cost to evaluate the quality of task partitioning.
The objective can be expressed as follows:}
\begin{equation}
\label{eq:EDP-core-assignment}
\begin{split}
&Minimize\quad O=\\
&\sum_{x}(r \cdot h_{x}+d_{x})\cdot cr_{x}+\varphi \cdot \sum_{x}{\sum_{y, y> x}{(r\cdot h_{xy}+d_{xy})\cdot cr_{xy}}}
\end{split}
\end{equation}
where \revise{$h_{x}$} and \revise{$d_{x}$} are average hop count and average Manhattan distance every two cores within community \revise{$cm_x$}, respectively, \revise{$cr_{x}$} is total communication requirements within community \revise{$cm_x$}, \revise{$h_{xy}$} and \revise{$d_{xy}$} are average hop count and average Manhattan distance every two hubs between community \revise{$cm_x$} and \revise{$cm_y$}, and \revise{$cr_{xy}$} represents communication requirements between these two community.
A penalty factor $\varphi$ ($>1$) of costs between communities allows source-sink pairs to be placed within the same community as much as possible.
After this stage, all tasks are initially assigned into communities.

\subsubsection{Placement in Communities}
\majrev{According to the tasks-to-community assignment, a detailed task placement method is proposed to place each task one by one into a specific core of the community to which it is assigned.
We adopt a greedy strategy to place tasks on the core with the least communication cost, and we use the hubs, which play a global key role in topology, to evaluate the communication cost between the current task and the unmapped tasks to further improve the quality of the mapping.}

\majrev{The task with the largest amount of communication is processed first, and it is placed on the hub node with the largest degree in the community to which it has been assigned.}
Subsequently, every time an unprocessed task that communicates the most with the processed tasks is selected for the next mapping. To minimize the communication cost, the task is placed on each unoccupied available core of the community to which it is assigned, and the communication cost is calculated, the task is placed on the core with the lowest communication cost.
In particular, for some tasks that are not placed to the specific cores but do communicate with the task being processed, the communication cost between the current task and the unmapped tasks cannot be calculated, but it affects the quality of the placement.
Note that for the sake of more accurate placement, the hubs in the community where the unprocessed tasks are allocated assume all its traffics, that is, simply place the task on these hubs.

During the detailed task placement, if there is currently maximum traffics between task \revise{$t_u$} and the tasks that has been placed, and \revise{$t_u$} is assigned to community \revise{$cm_x$}, we will try to place \revise{$t_u$} on each core \revise{$s_a\in cm_x$} and then calculate the communication cost \revise{$cost_{uv}^{s_a}$ of all flows in $G_t(T,E_t)$}.
The objective can be expressed as
\begin{subequations}
\label{eq:mapping}
\begin{align}
&\revise{Minimize\quad \sum_{t_v:(t_u,t_v)\, or\, (t_v,t_u)\in E_t}{cost_{uv}^{s_a}}} \\
&\revise{cost_{uv}^{s_a}=}
    \left\{
    \begin{array}{ll}
        (r\cdot h_{ab}+d_{ab})\cdot cr_{uv},   &\mbox{if}~ t_v\, has\, been\, placed;\\
        (r\cdot H_{ax}+D_{ax})\cdot cr_{uv}.   &\mbox{otherwise}.\\
    \end{array} \right.\nonumber\\
\end{align}
\end{subequations}
\revise{where $cr_{uv}$ is the communication requirement of flow $(t_u,t_v)$ in $E_t$, if $t_v$ has been placed in core $s_b$, $h_{ab}$ and $d_{ab}$ are respectively the hop count and Manhattan distance between core $s_a$ and $s_b$.
Otherwise, $H_{ax}$ and $D_{ax}$ indicate respectively the average hop count and average Manhattan distance between core $s_a$ and all hubs $hb_z$ of the community $cm_x$ where $c_b$ is assigned, respectively. The preliminary routing flow paths are allocated by Dijkstra's shortest-path algorithm.
If there is a minimum communication cost when $t_u$ is placed on $s_a$, $s_a$ is marked as unavailable. Repeat until all tasks are processed.}

\revise{Finally, according to the task-to-core mapping solution, we construct a core communication graph \majrev{$G_c$}: \majrev{$G(C,E_c)$} is a directed graph, where $\majrev{C}=\{c_x|1\leq x\leq |S|,c_x$ represents a core or switch\}, and \majrev{$E_c=\{(c_x,c_y)|$} there is a traffic flow $(c_x,c_y)\in E_c$ for each flow ($t_u,t_v$) $\in$ $E_t$ if task $t_u$ and $t_v$ are respectively assigned to core $s_x$ and $s_y$\}.}

\subsection{Deterministic Routing}
\label{sec:routing}
\revise{\noindent\textbf{Problem statement:} Given a  core communication graph \majrev{$G_c(C,E_c)$} and a \revise{brain-network-inspired} topology $G_s(S,E_s)$, we attempt to route all $|E_c|$ flows with minimization of the power consumption and hop count under the following constraints:}
\begin{itemize}
  \item \revise{the \textit{hop count}  constraint $LC^{xy}$ for each communication flow $(c_x,c_y)\in G_c(C,E_c)$,}
  \item \revise{and the \textit{bandwidth} constraint $cr_{cap}$ for physical links of $G_s(S,E_s)$.}
  \end{itemize}

We propose a Lagrangian relaxation-based \majrev{deadlock-free} routing algorithm to generate deterministic routing tables offline.
The routing problem as a multicommodity flow problem is an NP-hard problem \cite{book-Networkflow}, in which all individual commodities share a common facility.
\majrev{Therefore, to find a high-quality solution, rather than decomposing it into independent single-commodity flow problems to allocate routing paths one by one, it is more preferable to coordinate all communication flows to do path allocation \cite{book-Networkflow}.}

\majrev{With wormhole flow control, deadlocks can happen during routing of packets due to cyclic dependencies of resources (such as buffers) \cite{Huang2018Lagrangian,TurnProhibit}.
Given a network topology, we preprocess to break such cyclic dependencies by prohibiting certain turns to guarantee deadlock-free packet routing.
We use the turn prohibition (TP) algorithm presented in \cite{TurnProhibit,TurnProhibit2} to find the set of turns $PTS$ that need to be prohibited to break cycles, and at most 1/3 of all turns would be prohibited. When allocating the routing path, the paths are not allowed to go through the turns prohibited in $PTS$. }

For \majrev{deadlock-free} routing on the \revise{brain-network-inspired} NoC, minimizing communication power consumption and hop count is the optimization goal.
\revise{In $G_s(S,E_s)$, vertex $s_u\in S$ denotes a switch node and edge $(s_u,s_v)\in E_s$ represents a link that connects $s_u$ to $s_v$.
Let $t_{uv}^{xy}$ and $p_{uv}^{xy}$ represent flow $(c_x,c_y)$ across $(s_u,s_v)$ and its communication power, respectively. $p_{uv}^{xy}$ is defined as}
\begin{equation}
\label{eq:linkpower}
p_{uv}^{xy}= (J_{s_v} + J_{l_{uv}}) \cdot cr_{xy}
\end{equation}
where $cr_{xy}$ denotes the communication requirement of flow $(c_x,c_y)$, $J_{s_v}$ and $J_{l_{uv}}$ point the energy consumed by the switch $s_v$ and link $(s_u,s_v)$ for sending one bit of data, respectively.
Communication hop count is introduced into the objective function as a penalty. $\xi$ is a constant as the hop penalty factor for each link.
Then, the routing path allocation problem can be modeled \revise{as an ILP formulation}:
\begin{subequations}
\begin{align}
\label{eq:routing}
&Minimize~\sum_{(c_x,c_y)\in E_c}\sum_{(s_u,s_v)\in E_s}({p_{uv}^{xy}+\xi)\cdot t_{ux}^{xy}} \\
\textrm{s.t.} \nonumber\\
&\revise{Unit\ flows:}\sum_{s_v:(s_u,s_v)\in E_s}{t_{uv}^{xy}}-\sum_{s_v:(s_v,s_u)\in E_s}{t_{uv}^{xy}}=\nonumber\\
  &\quad\quad\quad\quad\quad\quad\quad\left\{
    \begin{array}{ll}
        1   &\mbox{if}~ s_u= s_k;\\
        0   &\mbox{if}~ s_u\in S-(s_k,d_k);\\
        -1   &\mbox{if}~ s_u= d_k;\\
    \end{array} \right.\label{eq:consflow}\\
&Hop\ count\ constraints:\nonumber\\
&\quad\quad\quad\sum_{(s_u,s_v)\in E_s}{t_{uv}^{xy}} \leq LC^{xy},\quad \forall (c_x,c_y)\in E_c; \\
&Bandwidth\ constraints:\nonumber\\
&\quad\quad\quad\sum_{(c_x,c_y)\in E_c}{cr_{uv}^{xy}\cdot t_{uv}^{xy}}\leq cr_{cap},\quad \forall (s_u,s_v)\in E_s;\\
&\quad t_{uv}^{xy}=0\ or\ 1.\label{eq:consbasic}
\end{align}
\end{subequations}

\revise{For large-scale flows, the ILP based method is very time-consuming.}
We propose a Lagrangian relaxation-based method to solve the multi-commodity flow problem.
\revise{Lagrangian relaxation is a solution technique that incorporates hard constraints of bandwidth and hop count into the objective using Lagrange multipliers and punishes the objective if they are not satisfied.}
The original problem is transformed to the Lagrangian subproblem:
\begin{subequations}
\label{eq:lagrangianRel}
\begin{align}
Min~&\sum_{(c_x,c_y)\in E_c}\sum_{(s_u,s_v)\in E_s}({p_{uv}^{xy}+\xi+\mu_{uv}\cdot cr_{uv}^{xy}+\mu^{xy})\cdot t_{ux}^{xy}}\nonumber\\
&-\sum_{(s_u,s_v)\in E_s}{\mu_{uv}\cdot cr_{cap}}-\sum_{(c_x,c_y)\in E_c}{\mu^{xy}\cdot LC^{xy}}\\
\textrm{s.t.}
&\quad (\ref{eq:consflow})\ and\ (\ref{eq:consbasic}) .
\end{align}
\end{subequations}

Since none of the constraints in this problem contains the flow variables for more than one of the commodities, the problem decomposes into separate \majrev{least-cost} path problems, one for each commodity.
\majrev{The problems are done by applying Dijkstra's shortest path algorithm \cite{book-Networkflow}, and only those paths that have turns not prohibited by $PTS$ can be selected.}
Then we solve the Lagrangian multiplier problem by using subgradient optimization \cite{Huang2018Lagrangian}.

\subsection{Complexity Analysis}
\majrev{Task mapping can be done in $O(N_M\cdot logN_M+|S|^2/N_M+(|E|+|S|\cdot log|S|))$ $(= O(|S|^2+|E|))$, where $N_M$ represents the number of communities (parts) acquired at the community detection stage.
The TP algorithm has been proven to be able to complete in $O(|S|^2\cdot m_a)$ \cite{TurnProhibit}.
The time complexity of counting the shortest routing path of $|E_c|$ flows in $G_c(C,E_c)$ is $O(|E_c|\cdot[|E|+|S|\cdot log|S|])$.}

\section{Experiments}
\label{sec:experiment}

The proposed synthesis method has been implemented in the C++ language and run on a Linux 64-bit workstation with Intel 2.0 GHz CPU and 64 GB memory.
\revise{We first verify the performance of the brain-network-inspired topology generated by the proposed method, i.e. the average hop count, power consumption, etc., and then verify the application mapping method with large-scale applications including real-world communication networks and synthetic applications.}
\revise{We use conventional mesh and torus as well as some irregular large-scale topologies for comparison experiments.
Since the average node degree of large-scale mesh topology is around 4, 
for a fair comparison, we assume that both of them use the same average degree of switches, that is, the same total number of links, and the initial degree of each new node $k$ is 2 and two new physical links are established.}

\subsection{Experimental Setup}
\subsubsection{Configuration of Our Model}
\label{sec:experiment-config}
In the experiments, we set the operating frequency at \majrev{1 GHz} and the data width for the NoC links (flit size) as 32 bits.
The switches for this validation have $2$ virtual channels (VCs) on each port. Each virtual channel can hold up to \majrev{4} flits.
The power model \cite{ICS} is used to estimate the power dissipation of the switches and physical links under 32-nm technology, whose technology parameters were extracted from the process model of the International Technology Roadmap for Semiconductors (ITRS) \cite{ITRS}.
We use the model \cite{ICS} to search for the optimal (size and number of) repeaters for each link and to calculate the power consumption.

\subsubsection{Testcase.}
\label{sec:experiment-bench}
\revise{Two real-world communication networks and synthetic applications are used to evaluate the proposed application mapping method.}
A real large network dataset collection \cite{snapnets} is provided by Stanford for large-scale graph processing.
A network with ground-truth communities, $email$-$Eu$-$core$, and Internet network, $p2p$-$Gnutella08$, are used in the performance evaluation of NoC designs.
\majrev{For graph processing in NoC-based distributed systems, such as in-memory computing architectures \cite{Neurocube,PIM-DATE,2014Neurogrid},
the large-scale graph data} is mapped to (stored in) many cores (vaults), iterating the vertices or edges of the graph in the current vault may require access to its adjacent vertices or edges in other vaults. Then application traffic \majrev{is} extracted from memory access dependencies between vaults caused by the scatter-gather operation \cite{2012PowerGraph}.
Benchmark $G\_3700$\ \&\ $G\_4096$ are constructed by combining the task graphs generated by Task Graphs For Free \cite{Huang2018Lagrangian}.
\majrev{The NoC architecture can also solve the interconnection communication between neurosynaptic cores as the basic processors.}
$VGG16$'s first two \majrev{fully} connected layers \cite{VGG} is abstracted as a benchmark $VGG16$, where network pruning is first performed to remove the trivial connections between neurons, and these synapses are partitioned into several fixed-size clusters. Between each partition, the number of connections translates into communication requirements.
In addition, \majrev{the article \cite{ChenGeneralized} provides benchmarks containing 30 to 70 tasks.} We merge all benchmarks and repeat them 20 times to form a large-scale case ($D\_3980$) \majrev{with relatively local communication, which is different from the above global communication cases.}
Table \ref{tab:bench} shows the detailed parameters of these benchmarks. Besides, Max-BW and Min-BW are the maximum and minimum communication requirements, respectively.
\begin{table}[htbp]
\centering\small
  \caption{The detailed parameters of all benchmarks.}
  \label{tab:bench}
  \renewcommand\tabcolsep{3.0pt}
  \begin{tabular}{ccccc}
    \toprule
    \textbf{Benchmarks}&\textbf{Task\ count}& \textbf{Flow\ count}& \textbf{Max-BW}& \textbf{Min-BW}\\
    \midrule
    $G\_3700$ & 3,700 & 6,301&100& 10\\
    $G\_4096$ & 4,096 & 7,108& 100& 10\\
    $D\_3980$& 3,980 &  5,300& 300& 2\\
    $VGG16$& 4,096 &  10,308& 64& 2\\
    $email$-$Eu$-$core$& 1,005 &  25,571& 15& 1\\
    $p2p$-$Gnutella08$& 6,301 &  20,777& 30& 1\\
  \bottomrule
\end{tabular}
\end{table}

\subsubsection{Simulation Configuration of $BookSim2$}
\label{sec:experiment-booksim}
A detailed and flexible cycle-accurate simulator $BookSim2$ \cite{Jiang2013A} is used to verify communication architecture performance, and it has been integrated with the power model \cite{ICS} to calculate the total network power consumption.
$BookSim2$ features a modular design and offers a set of configurable network parameters in terms of topology, routing algorithm, flow control, and switch microarchitecture.
Two extra routing algorithms, bandwidth-sensitive oblivious routing algorithm ($BSOR$) \cite{BSOR} and Lagrangian relaxation-based routing algorithm are also introduced into the tool.
We set all $LC_{k}$ and all $f_{cap}^{ij}$ to 12 and $4000MB/s$, respectively.

We present a detailed evaluation of \revise{brain-network-inspired} topology (BNIT), torus, and mesh for power and performance using four synthetic traffic patterns \cite{book:Interconnection,Jiang2013A}, including uniform, shuffle, bitcomp, and randperm.
The simulator's cycle time is a flit cycle, and the injection rate is specified in average packets per flit cycle per node.
However, it does not provide a way to simulate a real application.
To further verify routing algorithms and the performance of these communication architectures of which tasks have been mapped into cores, we operate a new customizable traffic pattern by mapping its communication graph into a customized deterministic table-based traffic.
We transform the communication requirement \majrev{$cr_{xy}$} ($MB/s$) of flow \majrev{$(r_x,r_y)\in G_r(R,E_r)$} into the packet injection rate $inject\_rate_{\majrev{xy}}\in (0\sim1)$ of that. $packet\_size$ denotes the number of flits per packet. \majrev{The packet injection rate} is expressed by
\begin{equation}
\label{eq:injection_rate}
inject\_rate_{\majrev{xy}} = \frac{cr_{xy}}{cr_{cap}\cdot packet\_size}
\end{equation}
Such table-based traffic allows specifying the source and destination pairs of packets along with the packet injection rate of each flow.

The complete list of configuration parameters used in the setup is summarized in Table \ref{tab:network-config}.
\majrev{In the synthetic traffic pattern and the application simulation, the packet size is 5 and 10 flits, respectively.}
More specifically, the transmission delays of each pipeline stage of switches and physical links are directly configured in the simulator.
As technology \majrev{scales}, global link delays due to wiring parasitics tend to dominate over gate delays in VLSI, which may make long-range links in NoC severely constrained by the link delay.
When repeaters are judiciously employed, the delay of transmitting signals on \majrev{the} global link can be reduced effectively by more than an order of magnitude \cite{ITRS,2015Networks}.
In our model, we assume that the distance between the adjacent switches is 0.1 $mm$.
\majrev{The delay (in $ns$) of physical links is extracted using RC delay models from \cite{ICS}.}

\begin{table*}[htbp]
\centering
\footnotesize
  \caption{Configuration parameters for simulation evaluation.}
  \label{tab:network-config}
  \renewcommand\tabcolsep{2pt}
  \begin{tabular}{ccc}
    \toprule
    \multicolumn{2}{c}{\centering{$\textbf{Configuration Parameter}$}} &{\centering{$\textbf{Value}$}}\\
    \midrule
    \multirow{6}{*}{\textbf{Network}}&Topology & BNIT\ \&\ Torus\ \&\ Mesh \\
    &Synthetic\ Traffic& Uniform\ \&\ Shuffle \&\ Bitcomp \&\ Randperm\\
    &Operating Frequency& \majrev{1}\ GHz\\
    &Warm-up\ Period&\majrev{$30k$}\ $cycles$\\
    &Simulation\ Period&\majrev{$100k$}\ $cycles$\\
    &Packet\ Size& \majrev{5 or 10} flits\\
    &\majrev{Flit\ Size}& \majrev{32 bits}\\
    \midrule
    \multirow{5}{*}{\textbf{Switch}}&Switch\ Type&Input-queued Architecture\\
    &Flow\ Control &Wormhole\\
    &Number\ of\ VCs per Port&2\\
    &VC\ Buffer\ Size&\majrev{4} flits\\
    &Switch Pipeline Stages&4\\
    &VC Allocator/Switch Allocator/Routing Delay&1 $cycle$\\
    \bottomrule
\end{tabular}
\end{table*}

\subsection{Topology Analysis}
\label{sec:exp-result}
\subsubsection{Effect of $\gamma$ and $\beta$}

\begin{table*}[htbp]
\centering\small
  \caption{The value of $m_a$ when $\gamma$ is varied from 0.5 to 2.5 and $m$ is 15.}
  \label{tab:ma}
  \renewcommand\tabcolsep{7.0pt}
  \begin{tabular}{|c|c|c|c|c|c|c|c|c|c|}
    \hline
    $\bm{\gamma}$&0.5 $\sim$ 0.6&0.7 $\sim$ 1.0&1.1 $\sim$ 1.3&1.4 $\sim$ 1.5&1.6&1.7&1.8&1.9&$\geq2.0$\\
    \hline
    $\bm{m_a}$&7&8&9&10&11&12&13&14&15\\
    \hline
\end{tabular}
\end{table*}

\begin{figure}[htbp]
\small \centering
  \includegraphics[width=6cm]{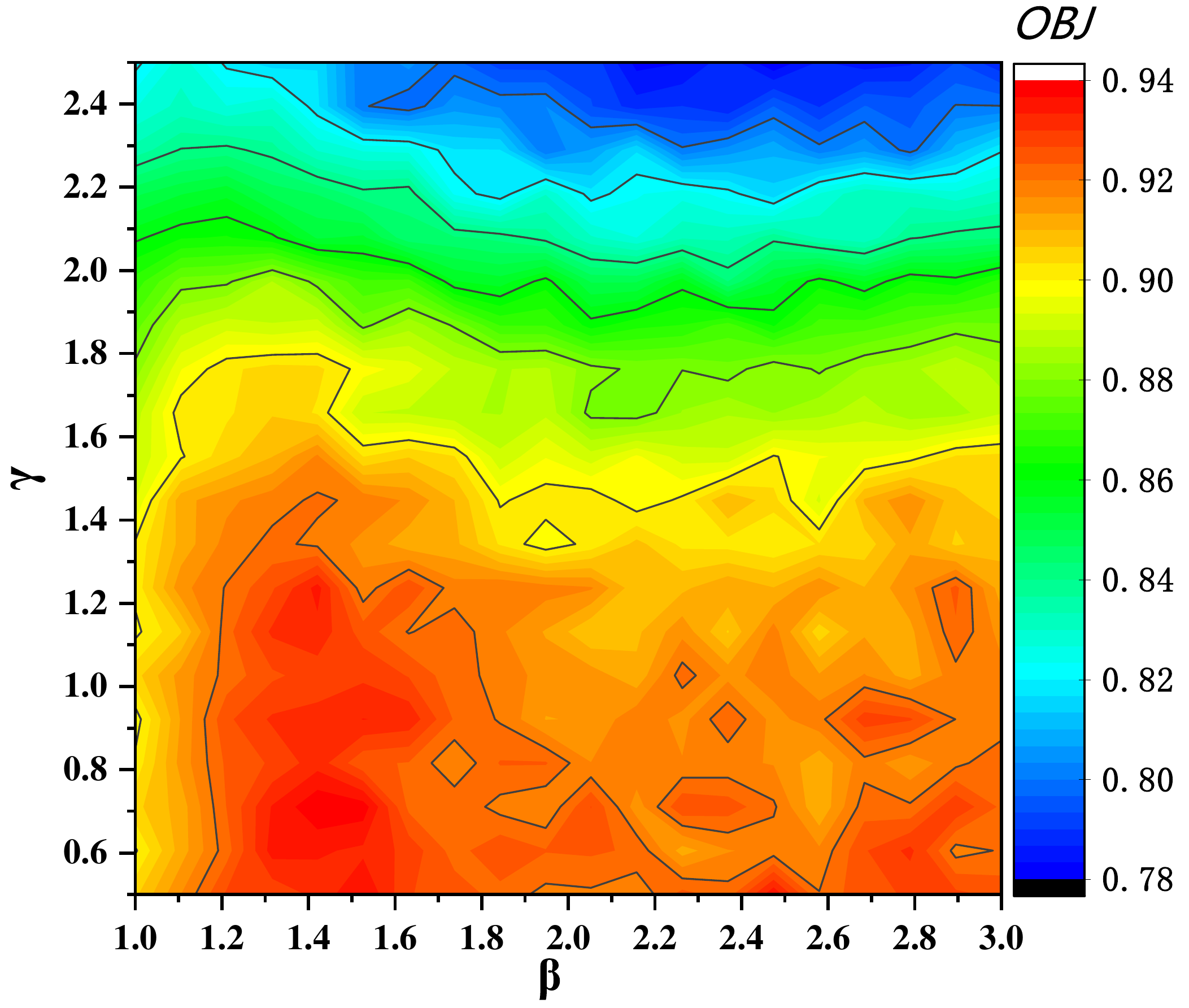}\\
  \caption{\revise{Variation of $OBJ$ w.r.t. exponent $\gamma$ \& $\beta$.}}
  \label{fig:exp}
\end{figure}

\begin{figure}[htbp]
\small \centering
  \includegraphics[width=5.2cm]{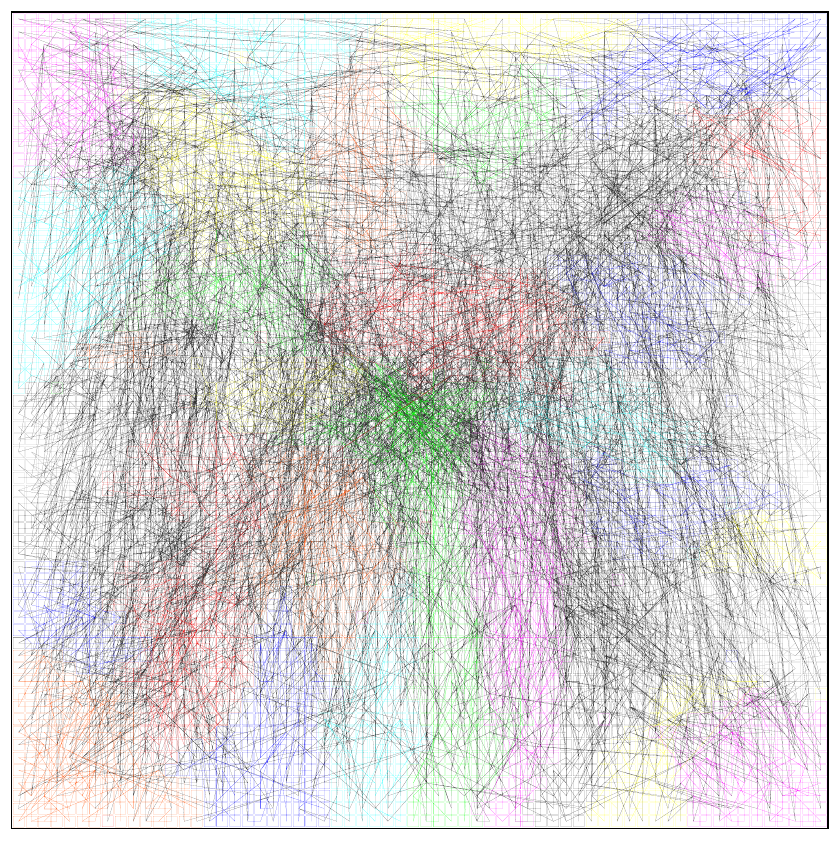}\\
  \caption{Final topology of 4096 (64*64) nodes. For easy identification, the colors of the regions and links represent the communities to which they belong.}
  \label{fig:finaltopo}
\end{figure}

\begin{figure}[htbp]
\small \centering
  \includegraphics[width=7.5cm]{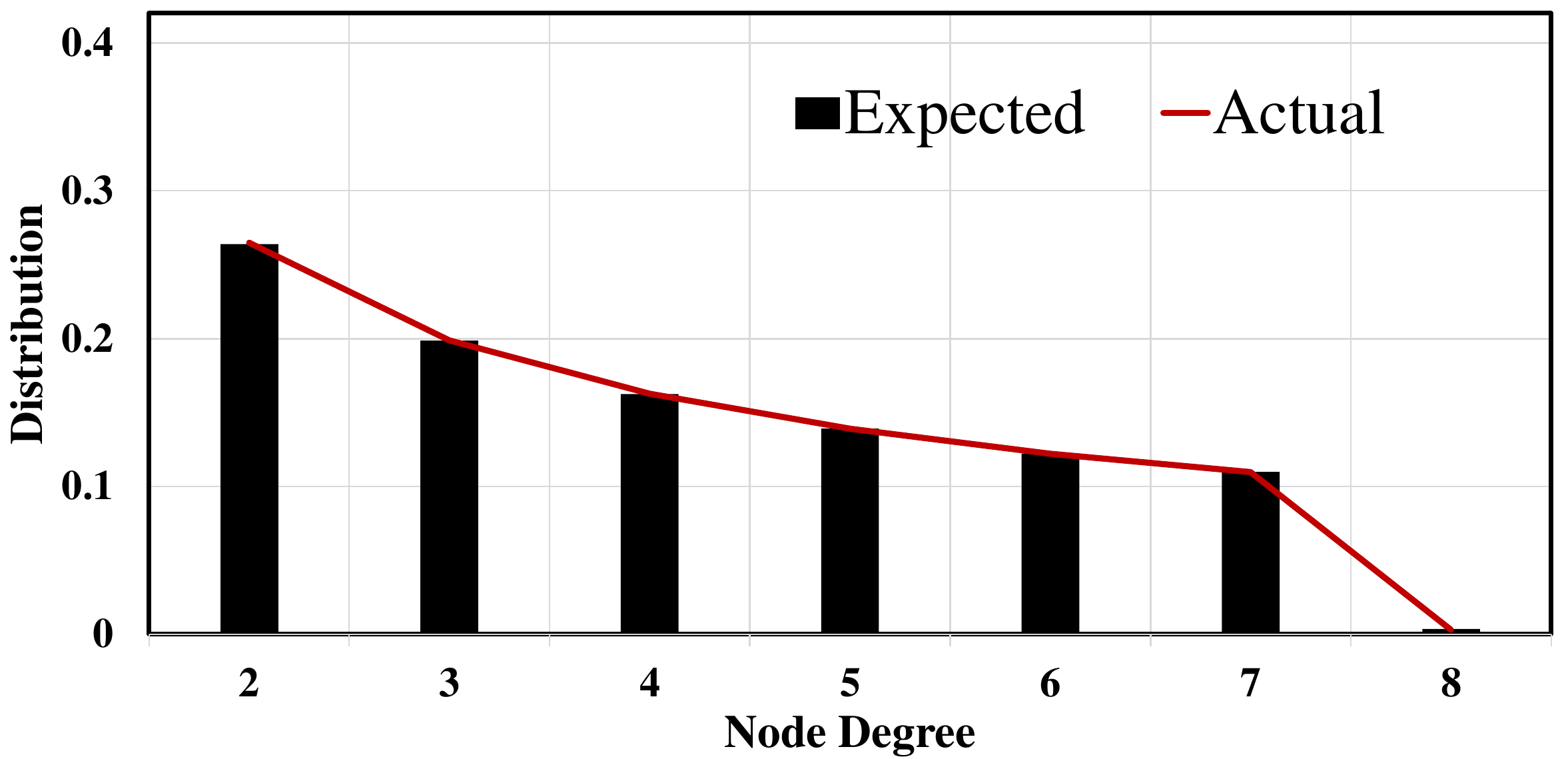}\\
  \caption{Degree distribution of nodes of BNIT with $\gamma=$ \majrev{0.7} and $n=4096$.}
  \label{fig:deg}
\end{figure}

\begin{figure}[htbp]
\small \centering
  \includegraphics[width=7.5cm]{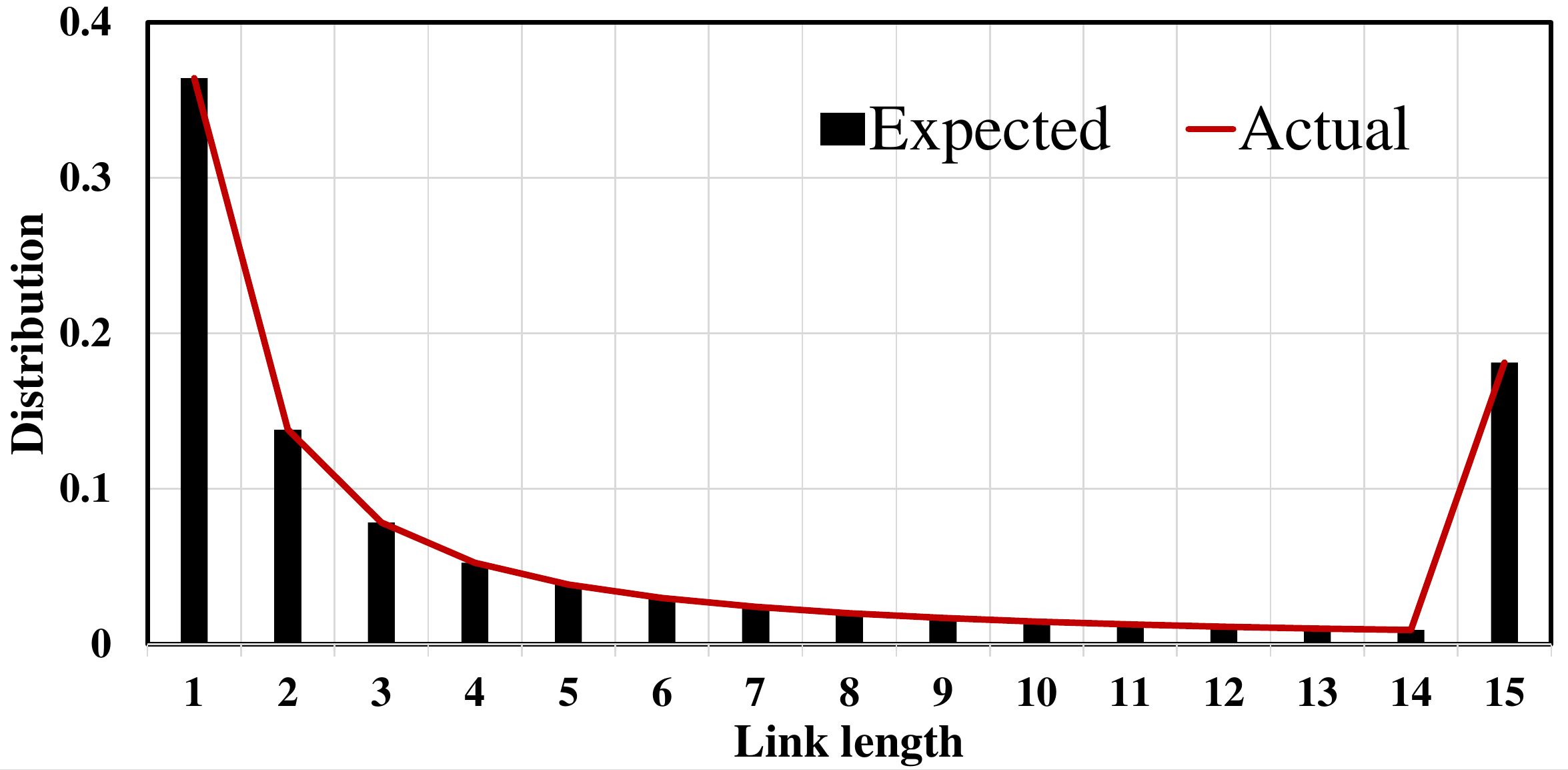}\\
  \caption{Link length distribution of BNIT with $\beta=$ \majrev{1.4}  and $n=4096$. The unit of link length is the distance between adjacent rows or columns in the coordinate.}
  \label{fig:leg}
\end{figure}

An example with $n=4096$ (same with TrueNorth \cite{TrueNorth2014}), \majrev{$l_a=15$}, and $m=15$ is presented.
We vary respectively $\beta$ from 1.0 to 3.0 and $\gamma$ from \majrev{0.5 to 2.5}, including those that occur most frequently in brain networks \cite{2008Small}.
\revise{In addition, to satisfy the inequality \ref{eq:ineq}, the value of $m_a$ is shown in Table \ref{tab:ma}.}
Figure \ref{fig:exp} shows the $OBJ$ value (equation \ref{eq:obj-topo}) of BNIT generated by the proposed method for each $\gamma$ and $\beta$.
We can see that when \majrev{$\gamma=0.7$} and \majrev{$\beta=1.4$}, the $OBJ$ has the maximum value, and \revise{a proper} network topology can be obtained \revise{and shown in Figure \ref{fig:finaltopo}}.
In Figure \ref{fig:deg}, we show the degree distribution of BNIT constructed by our algorithm. The actual degree distributions perfectly match with the expected degree distributions with \majrev{$\gamma=0.7$}.
Based on the preferential implementation of the scale-free property, we further make the edge length distribution tend to the power-law small-world property. Figure \ref{fig:leg} shows that the overall trend of the actual link length distribution and the expected link length distribution are close, even if there is a small deviation in some intervals.

\majrev{Subsequently, each link of brain-network-inspired topology has a specific length, and the length should be guaranteed during topology metal routing. An existing method, bounded-length maze routing algorithm \cite{BLMR}, can be used to solve the metal routing problem of this topology.}

\majrev{\subsubsection{Effect of $m$ and $l_a$}
$m$ and $l_a$ are respectively the user-defined maximal switch size and maximal link length.
As shown in Table \ref{tab:mla}, \textbf{\#hop} is the average minimum hop count between every two nodes.
$\bm{P}$ and $\bm{C}$ represent respectively the basic power consumption, \majrev{including static power and clocking power}, and rough communication cost.
The total link length \textbf{\#WL} is defined as the sum of the lengths of all links, and the unit of length is the distance between adjacent rows or columns.
When the topology generation is accompanied by larger $m$ and $l_a$, the generated network topology may have lower average hop count.}

\begin{table}[htbp]
\centering
\small
  \caption{Different $m$ and $l_a$ correspond to the parameters of the topology with 1024 nodes.}
  \label{tab:mla}
  \renewcommand\tabcolsep{10.0pt}
  \begin{tabular}{cccccc}
    \toprule
    $m$&$l_a$ &\textbf{\#hop}&$\bm{P}$&$\bm{C}$ &\textbf{\#WL}\\
    \midrule
			10&10&8.03&4.58&1.4E4&1233\\
			10&15&7.33&4.65&1.4E4&1503\\
			15&10&8.03&4.58&1.4E4&1233\\
			15&15&7.33&4.65&1.4E4&1503\\
			15&20&7.06&4.71&1.4E4&1754\\
			15&30&6.96&4.82&1.7E4&2212\\
    \bottomrule
\end{tabular}
\end{table}

\subsubsection{Comparison with Conventional Regular Topologies and Irregular Topologies}
\label{sec:topo-compair}
\begin{figure}[htbp]
\small \centering
\subfigure[]{
	\centering
	\includegraphics[height=4cm]{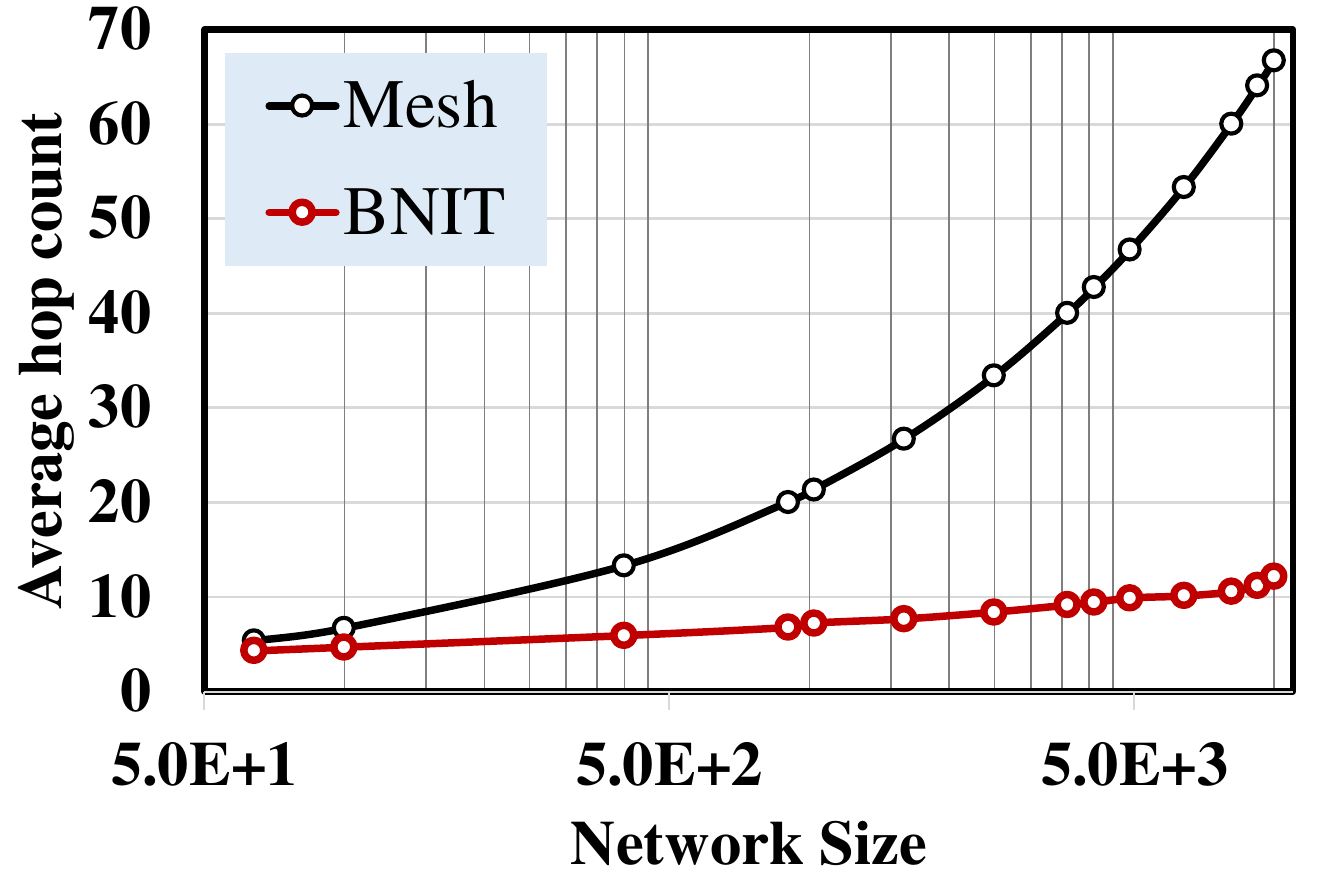}
	\label{fig:SW1}
}
\hspace{5ex}
\subfigure[]{
	\centering
	\includegraphics[height=4cm]{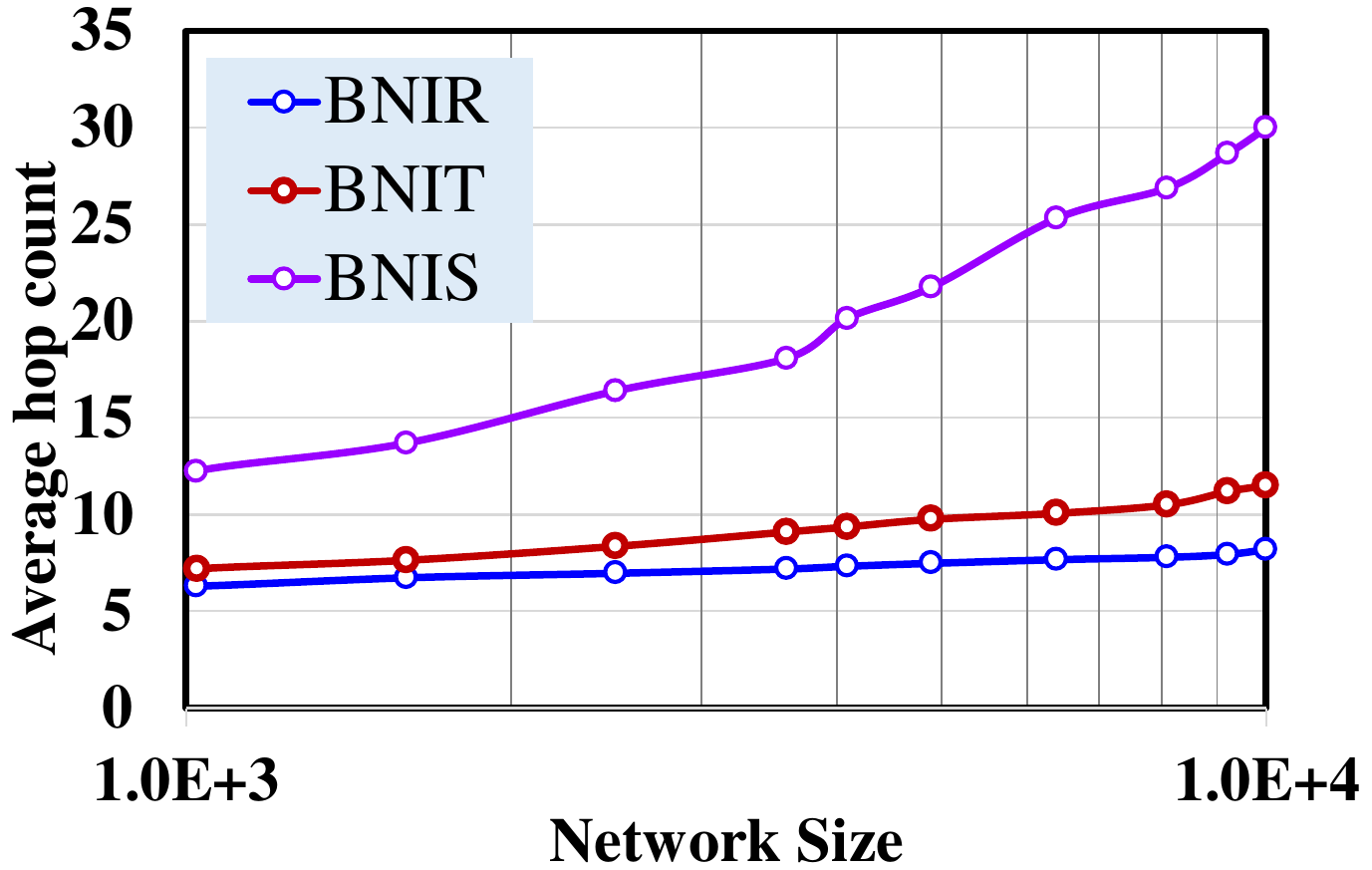}
	\label{fig:SW2}
}
  \caption{Scaling of the average hop counts as a function of the network size. (a) The contrast between mesh and BNIT. (b) The contrast between BNIR, BNIT, and BNIS.}
  \label{fig:SW}
\end{figure}

Compared to mesh, BNIT presents an extremely lower average hop count proportional to the logarithm of the network size (is called small-world), which Figure \ref{fig:SW1} confirms.
Meanwhile, when the network size is greater than 100, the average hop count of BNIT is 35\% to 90\% lower than that of Mesh.
When the network size is less than 100, although the two \majrev{are} very close, the average hop count of BNIT is still smaller than \majrev{the} mesh's.
So BNIT generated by the proposed method is more suitable for networks with \majrev{a} size larger than 100.

\begin{figure}[htbp]
\small \centering
  \includegraphics[width=6.5cm]{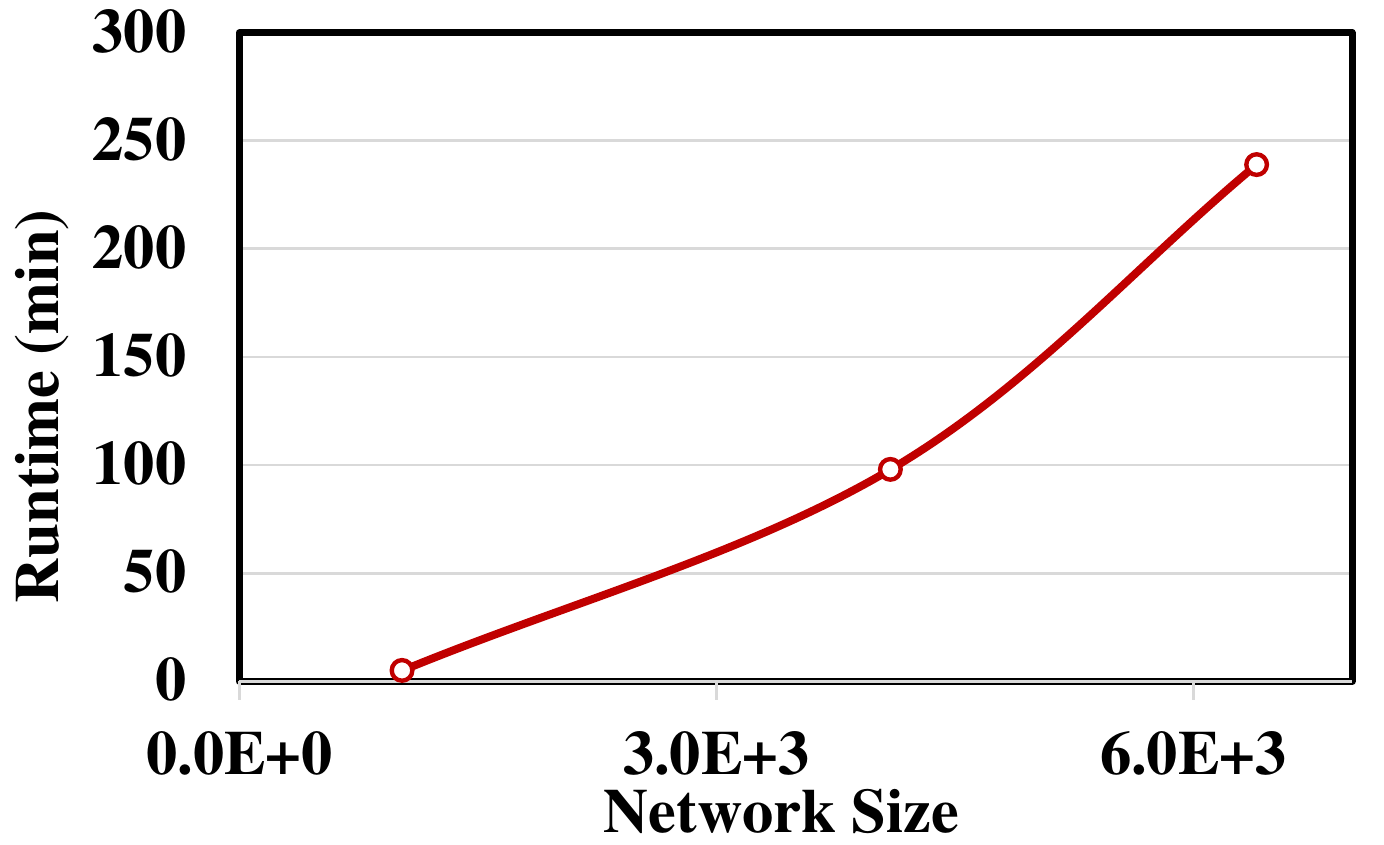}\\
  \caption{Runtime of topology generation and community detection. }
  \label{fig:runtime}
\end{figure}

In Figure \ref{fig:SW2}, legend BNIR and BNIS mean respectively that the random establishment without consideration of link length \revise{(SDA \cite{BulutConstructing}}) and the greedy selection of nodes with the shortest link length, rather than the pursuit of power-law distributions of link length in $Line$ $23\sim30$ of Algorithm \ref{alg:topogen}.
\majrev{Brain-network-inspired} topologies with power-law length distribution (BNIT) have \majrev{at least 50\%} lower average hop counts \majrev{than BNIS}.
The advantage of BNIT's low average hop count will be highlighted as the network size increases.
\majrev{The average hop count of BNIR is slightly lower than that of BNIT.
Figure \ref{fig:runtime} shows the runtime of the topology generation and community detection.}

BA network based on growth and preferential attachment \cite{Guclu2009Limited} is implemented to construct a scale-free peer-to-peer network with fixed exponent $\gamma$.
Here, we migrate this model to generate a scale-free BA topology for NoC.
As shown in Table \ref{tab:topo}, the experimental results were derived from different-scale conventional regular topologies and irregular complex topologies, and $n$ is set to 1024, 4096, 6400, and 9216, respectively.
\majrev{\textbf{\#link} is the number of links of topologies.}

Different from the mesh of $t\times t$, torus also has $2t$ long-range links across the plane in addition to short links of one unit length.
\majrev{The communication cost of torus is higher than that of mesh. This is because} the routing path allocation is based on the minimum hop count, many node pairs would choose the path through the crossed long-range links to reduce the hop count between \majrev{every} two nodes, which results in greater \majrev{communication cost}.
\begin{table*}[htbp]
\centering
  \caption{Comparison between mesh, torus, and BNIT with different network sizes.}
  \label{tab:topo}
  \footnotesize
  \renewcommand{\arraystretch}{1.3}
  \renewcommand\tabcolsep{1.5pt}
  \begin{tabular}{|c|c|cc|cc|cc|cc|cc|cc|}
    \hline
    $\textbf{Topology}$& $\textbf{Metric}$ &\multicolumn{2}{c|}{\centering{$\textbf{BNIT}$}} &\multicolumn{2}{c|}{\centering{$\textbf{BNIR (SDA)}$}} &\multicolumn{2}{c|}{\centering{$\textbf{BNIS}$}} & \multicolumn{2}{c|}{\centering{$\textbf{BA\ model}$}} &\multicolumn{2}{c|}{\centering{$\textbf{Torus}$}} &\multicolumn{2}{c|}{\centering{$\textbf{Mesh (base.)}$}} \\
    \hline
    \multirow{5}{*}{\textbf{$1024$}}
        &\textbf{\#hop}&\majrev{7.33}&\majrev{\textbf{-66\%}}&\majrev{6.32}&\majrev{-70\%}&\majrev{12.24}&\majrev{-43\%}&\majrev{5.59}&\majrev{-73\%}&17.02&\majrev{-20\%}&21.33&1\\
            \cline{2-14}
            &$\bm{C}$&\majrev{1.4E4}&\majrev{\textbf{-53\%}}&\majrev{3.0E4}&\majrev{0\%}&\majrev{2.1E4}&\majrev{-30\%}&\majrev{2.0E4}&\majrev{-30\%}&3.2E4&\majrev{+7\%}&3.0E4&1\\
                \cline{2-14}
                &$\bm{P}$&\majrev{4.65}&\majrev{\textbf{+19\%}}&\majrev{6.05}&\majrev{+54\%}&\majrev{4.49}&\majrev{+15\%}&\majrev{8.02}&\majrev{+69\%}&\majrev{4.03}&\majrev{+3\%}&\majrev{3.92}&1\\
                    \cline{2-14}
                   &$\textbf{\#link}$&2040&\textbf{+3\%}&2040&+3\%&2040&+3\%&2040&+3\%&2048&+3\%&1984&1\\
                       \cline{2-14}
                      &$\textbf{\#WL}$&\majrev{1503}&$\majrev{\bm{\times 3.8}}$&\majrev{7313}&\majrev{$\times 18.4$}&\majrev{755}&\majrev{$\times 1.9$}&\majrev{7048}&\majrev{$\times 18.0$}&794&\majrev{$\times 2.0$}&397&1\\
    \hline
    \hline
        \multirow{5}{*}{\textbf{$4096$}}
        &\textbf{\#hop}&\majrev{9.37}&\majrev{\textbf{-78\%}}&\majrev{7.34}&\majrev{-83\%}&\majrev{20.14}&\majrev{-53\%}&\majrev{6.46}&\majrev{-85\%}&33.00&\majrev{-23\%}&42.67&1\\
            \cline{2-14}
            &$\bm{C}$&\majrev{5.4E5}&\majrev{\textbf{-72\%}}&\majrev{1.3E6}&\majrev{-31\%}&\majrev{9.4E5}&\majrev{-51\%}&\majrev{8.4E5}&\majrev{-55\%}&2.0E6&\majrev{+5\%}&1.9E6&1\\
                \cline{2-14}
                &$\bm{P}$&\majrev{19.53}&\majrev{\textbf{+24\%}}&\majrev{31.69}&\majrev{+101\%}&\majrev{18.68}&\majrev{+19\%}&\majrev{40.03}&\majrev{+131\%}&\majrev{16.15}&\majrev{+3\%}&\majrev{15.73}&1\\
                    \cline{2-14}
                   &$\textbf{\#link}$&8184&\textbf{+1\%}&8184&+1\%&8184&+1\%&8184&+1\%&8192&+2\%&8064&1\\
                       \cline{2-14}
                      &$\textbf{\#WL}$&\majrev{8720}&$\majrev{\bm{\times  5.4}}$&\majrev{5.9E4}&\majrev{$\times 36.9$}&\majrev{3480}&\majrev{$\times 2.2$}&\majrev{5.7E4}&\majrev{$\times 35.0$}&3226&\majrev{$\times 2.0$}&1613&1\\
    \hline
    \hline
        \multirow{5}{*}{\textbf{$6400$}}
        &\textbf{\#hop}&\majrev{10.97}&\majrev{\textbf{-80\%}}&\majrev{7.68}&\majrev{-86\%}&\majrev{26.31}&\majrev{-51\%}&\majrev{6.71}&\majrev{-87\%}&40.30&\majrev{-24\%}&53.30&1\\
            \cline{2-14}
            &$\bm{C}$&\majrev{2.0E6}&\majrev{\textbf{-73\%}}&\majrev{2.3E6}&\majrev{-68\%}&\majrev{4.3E6}&\majrev{-41\%}&\majrev{2.7E6}&\majrev{-63\%}&7.7E6&\majrev{+5\%}&7.3E6&1\\
                \cline{2-14}
                &$\bm{P}$&\majrev{30.60}&\majrev{\textbf{+24\%}}&\majrev{55.09}&\majrev{+124\%}&\majrev{28.16}&\majrev{+15\%}&\majrev{66.95}&\majrev{+163\%}&\majrev{25.24}&\majrev{+3\%}&\majrev{24.59}&1\\
                    \cline{2-14}
                   &$\textbf{\#link}$&12792&\textbf{+1\%}&12792&+1\%&12792&+1\%&12792&+1\%&12800&+1\%&12640&1\\
                       \cline{2-14}
                      &$\textbf{\#WL}$&\majrev{1.4E4}&$\majrev{\bm{\times 5.5}}$&\majrev{1.2E5}&\majrev{$\times 45.8$}&\majrev{4696}&\majrev{$\times 1.9$}&\majrev{1.1E5}&\majrev{$\times 44.0$}&5056&\majrev{$\times 2.0$}&2528&1\\
    \hline
    \hline
        \multirow{5}{*}{\textbf{$9216$}}
        &\textbf{\#hop}&\majrev{11.35}&\majrev{\textbf{-82\%}}&\majrev{7.95}&\majrev{-88\%}&\majrev{30.66}&\majrev{-52\%}&\majrev{6.94}&\majrev{-89\%}&49.00&\majrev{-23\%}&64.00&1\\
            \cline{2-14}
            &$\bm{C}$&\majrev{4.7E6}&\majrev{\textbf{-78\%}}&\majrev{1.2E7}&\majrev{-46\%}&\majrev{1.3E7}&\majrev{-38\%}&\majrev{7.2E6}&\majrev{-65\%}&2.3E7&\majrev{+10\%}& 2.1E7&1\\
                \cline{2-14}
                &$\bm{P}$&\majrev{45.66}&\majrev{\textbf{+29\%}}&\majrev{87.48}&\majrev{+147\%}&\majrev{40.56}&\majrev{+14\%}&\majrev{104.4}&\majrev{+193\%}&\majrev{36.34}&\majrev{+3\%}&\majrev{35.43}&1\\
                    \cline{2-14}
                   &$\textbf{\#link}$&18424&\textbf{+1\%}&18424&+1\%&18424&+1\%&18424&+1\%&18432&+1\%&18240&1\\
                       \cline{2-14}
                      &$\textbf{\#WL}$&\majrev{2.3E4}&$\majrev{\bm{\times 6.3}}$&\majrev{2.0E5}&\majrev{$\times 54.8$}&\majrev{6762}&\majrev{$\times  1.9$}&\majrev{1.9E5}&\majrev{$\times 52.8$}&7296&\majrev{$\times 2.0$}&3648&1\\
    \hline
\end{tabular}
\end{table*}

Compared to mesh and torus, BNIT generated by the proposed method reduces the average hop count by about 80\%, and reduces rough \majrev{communication cost} by about 70\%, although the \majrev{basic} power is slightly increased by around 20\%.
As the network size increases, the average hop count of BNIT is getting lower and lower compared with mesh and torus.
The number of links of all topologies is indistinguishable. Since the initial degree of the nodes in BNIS, BINR, and BNIT is the same, they have the same number of links.
Compared with the topology (BNIR) generated by SDA, BNIT can reduce \majrev{communication cost and basic} power consumption by about \majrev{50\% and 40\%}, respectively, and by employing only a small fraction of the long-range links, BNIT has approximately \majrev{85}\% lower the total link length.
Compared with mesh, although the average hop count of BNIR is reduced by at least \majrev{70}\%, it will \majrev{result in an even doubling of the basic power consumption and a 50-fold increase in the total link length.
In addition, BNIR has fewer short-distance links, which may cause the communication between adjacent nodes to pass through some long-distance links.
Therefore, compared with BNIT, it requires more power consumption in link transmission, which increases communication cost.}
Due to the lack of a proper amount of long-range links, BNIS has a higher average hop count and rough \majrev{communiation} power consumption than BNIT, which is the same as shown in Figure \ref{fig:SW2}.
Despite the low average hop count and \majrev{communication cost}, the BA model \cite{OshidaPacket} is not flexible, such as the fixed exponent in \majrev{the} power-law degree distribution, no switch size constraint (high-radix switches), and no link length perception (many long-range links), and the \majrev{basic} power consumption of the generated topology is even up to 3 times that of mesh.
Therefore, this kind of inflexible generation of BA model is unaccommodated to NoC.

\begin{figure*}[htbp]
\centering
\subfigure[]{
	\centering
	\includegraphics[height=4.5cm]{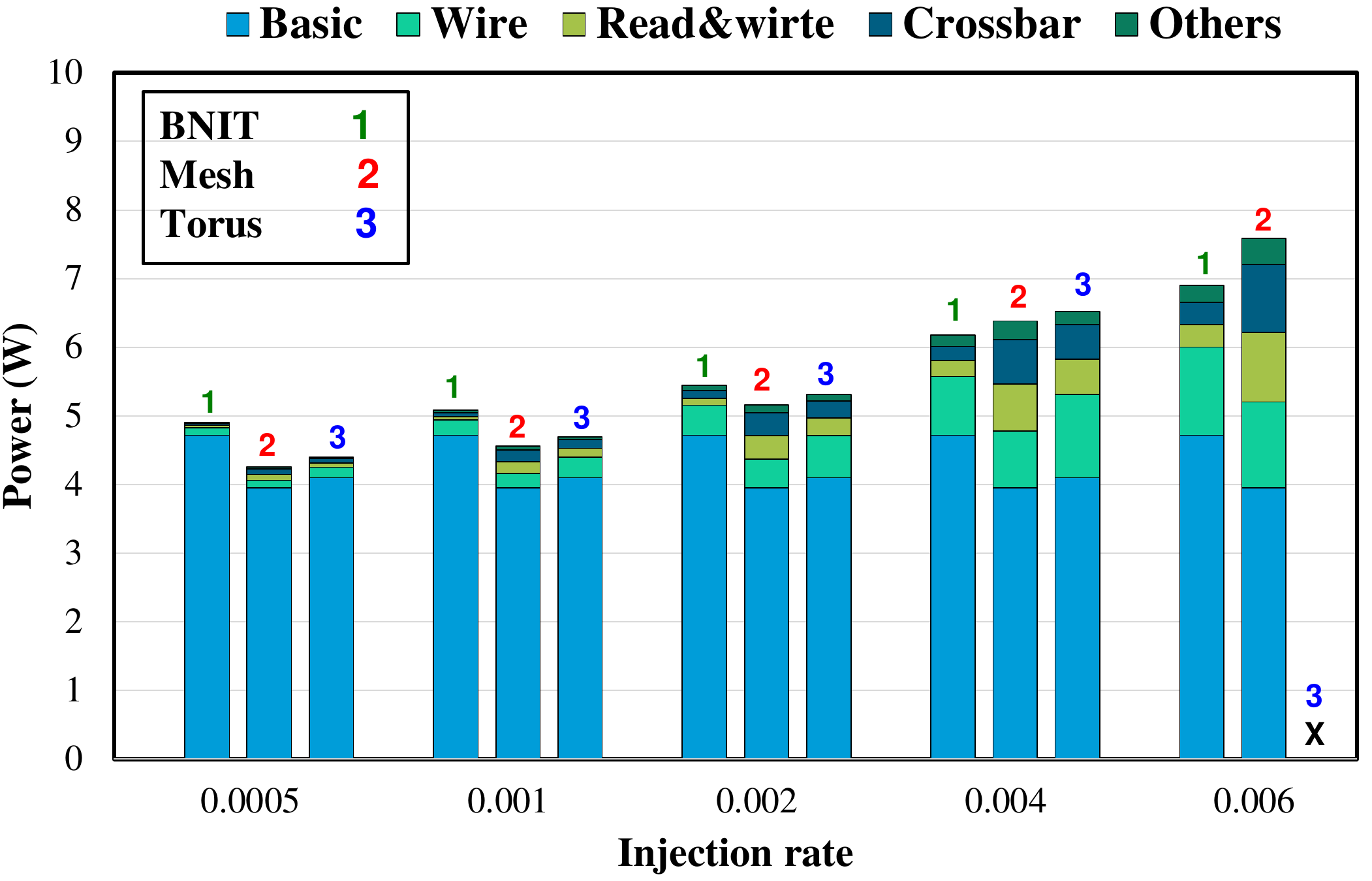}
	\label{fig:uniform}
}
\hspace{4ex}
\subfigure[]{
	\centering
	\includegraphics[height=4.5cm]{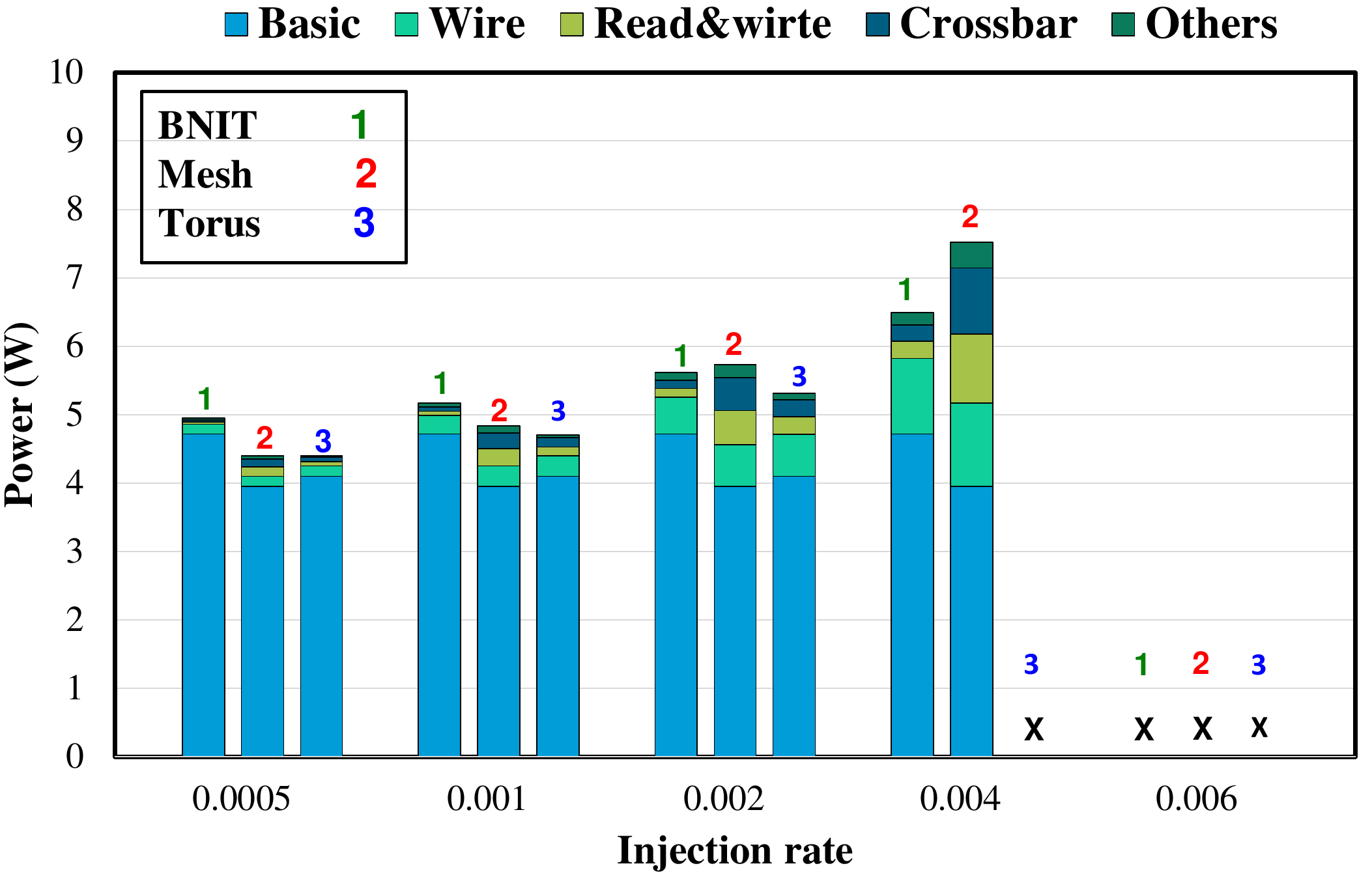}
	\label{fig:shuffle}
}\\
\subfigure[]{
	\centering
	\includegraphics[height=4.5cm]{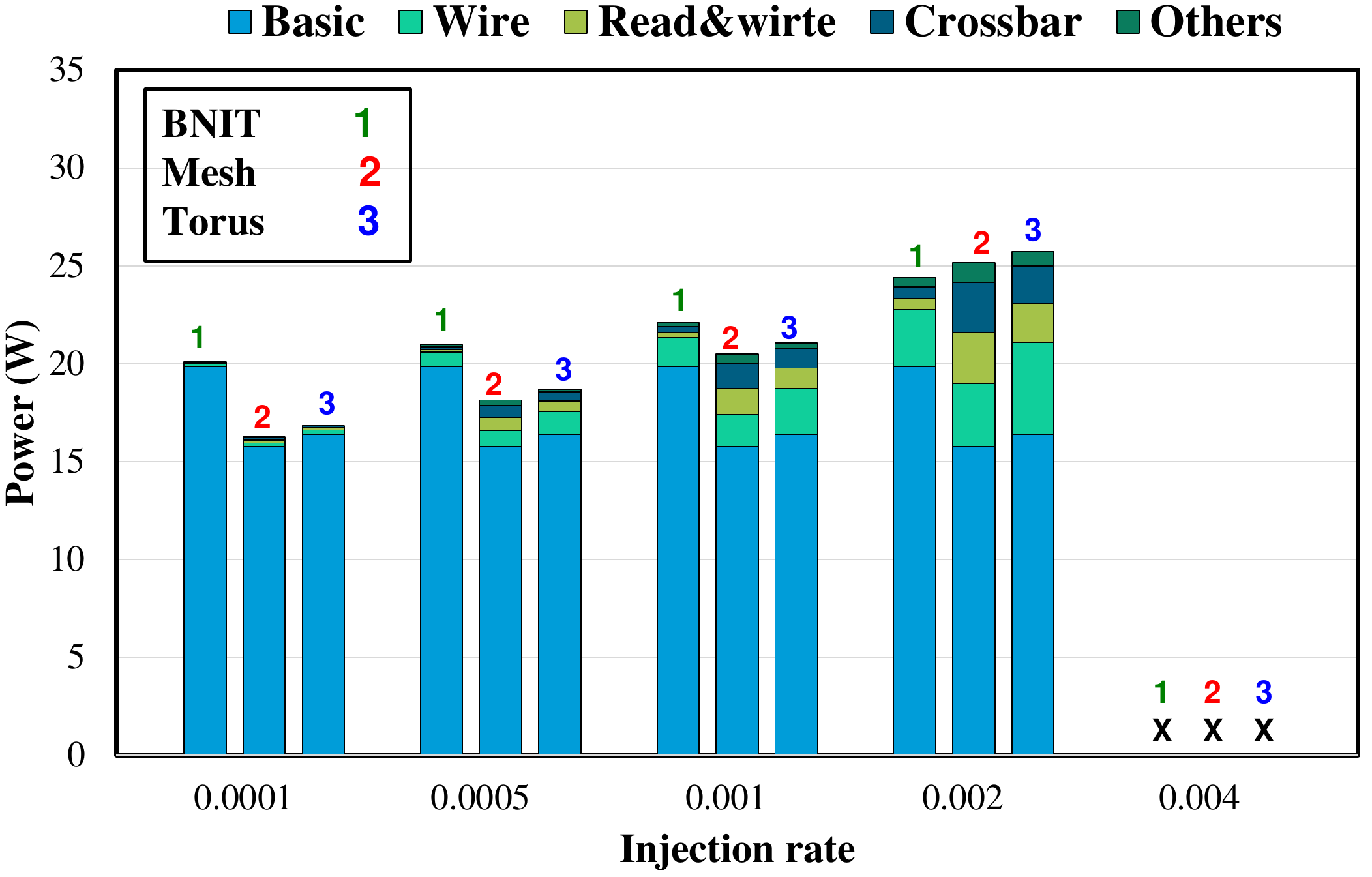}
	\label{fig:uniform}
}
\hspace{4ex}
\subfigure[]{
	\centering
	\includegraphics[height=4.5cm]{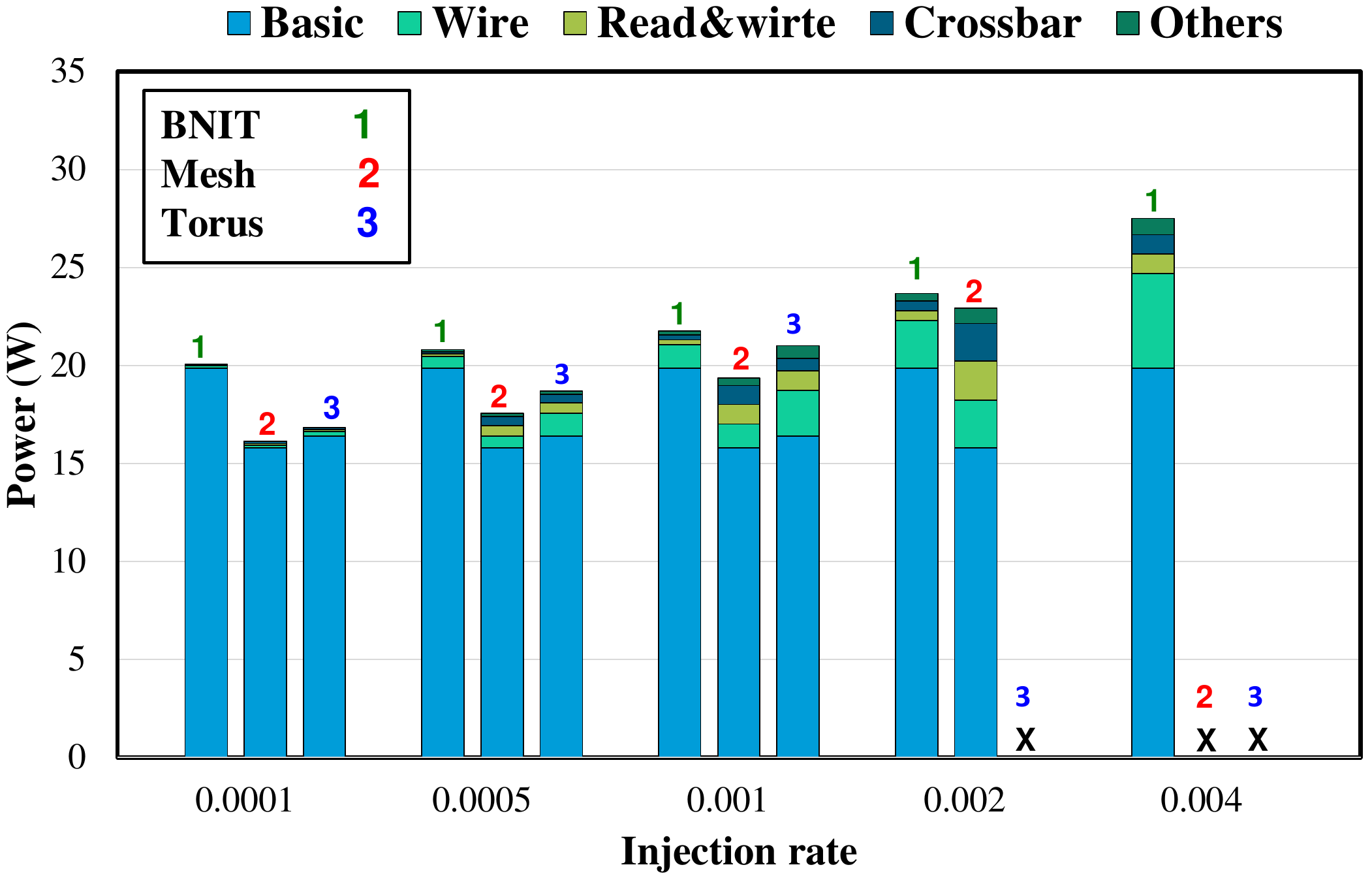}
	\label{fig:shuffle}
}\\
\subfigure[]{
	\centering
	\includegraphics[height=4.5cm]{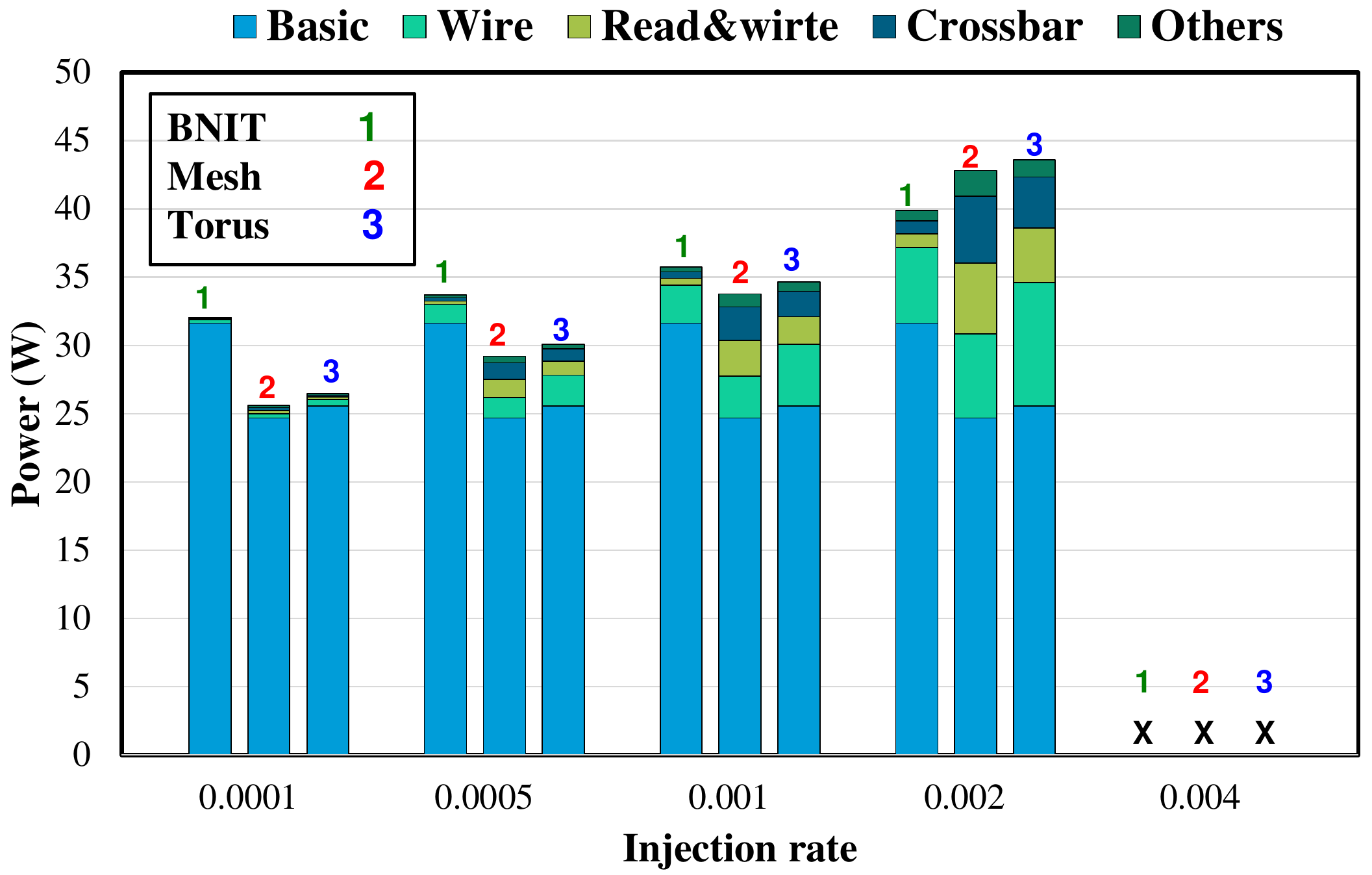}
	\label{fig:uniform}
}
\hspace{4ex}
\subfigure[]{
	\centering
	\includegraphics[height=4.5cm]{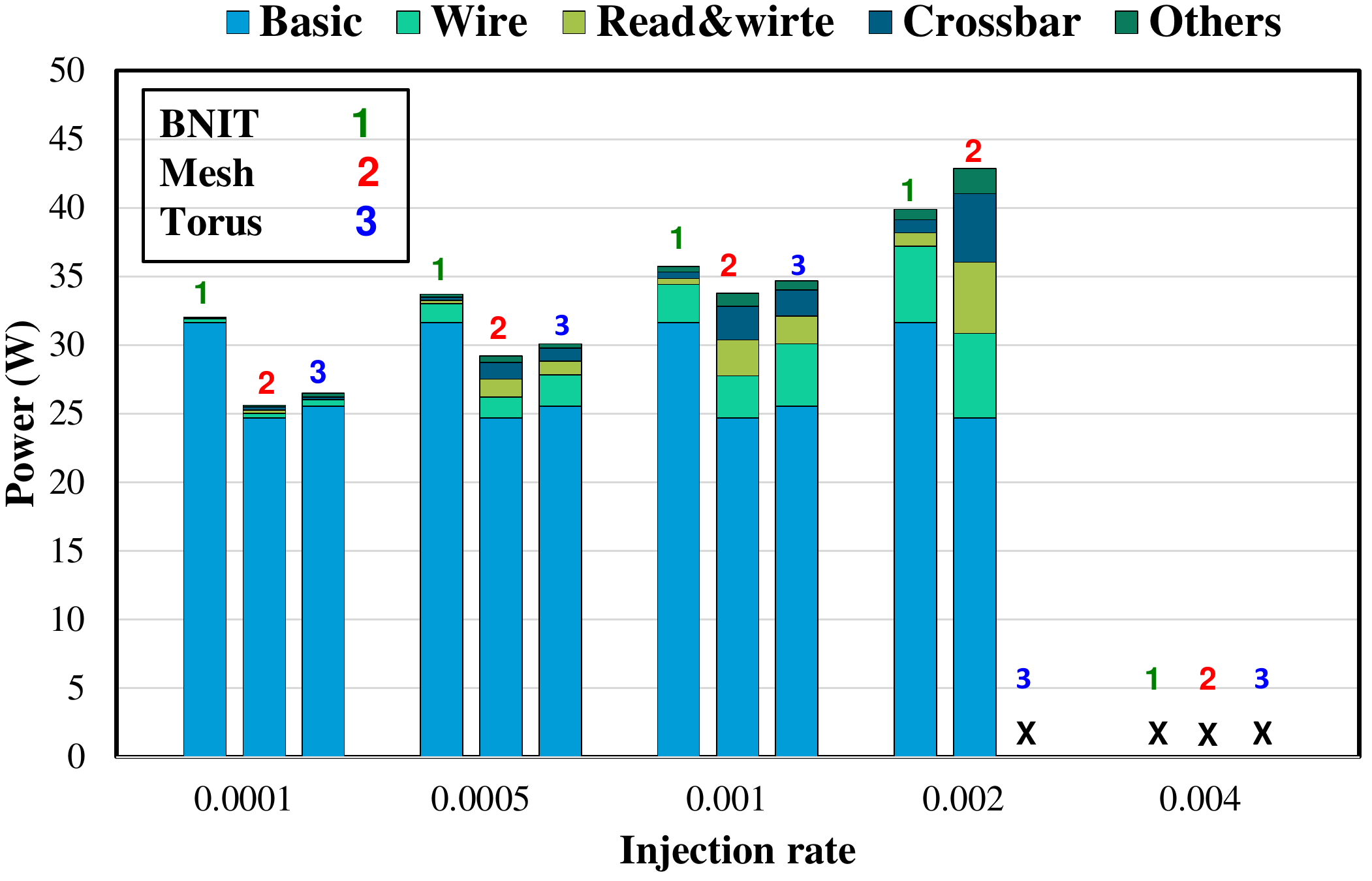}
	\label{fig:shuffle}
}
    \caption{\revise{Power evaluation for mesh, torus, and BNIT with 1024 cores at (a) uniform and (b) bitcomp traffic patterns, with 4096 cores at (c) uniform and (d) shuffle traffic patterns, and with 6400 cores at (e) uniform and (f) randperm traffic patterns.}}
    \label{fig:booksim1}
\end{figure*}
\subsubsection{\revise{Performance Validation under Different Traffic Patterns Using $BookSim2$}}
\label{sec:trafficpattern}

\begin{figure*}[htbp]
\centering
\subfigure[]{
	\centering
	\includegraphics[height=3cm]{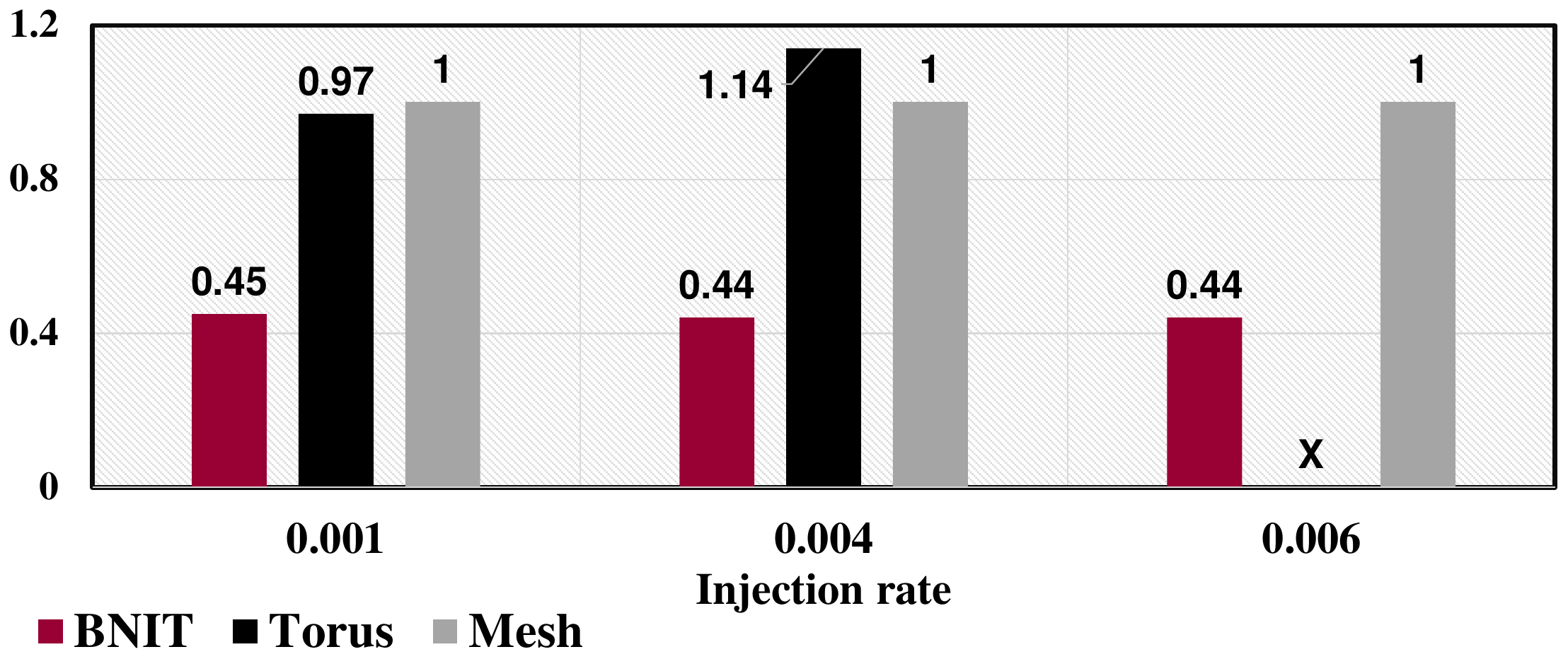}
	\label{fig:uniform-lan}
}
\hspace{5ex}
\subfigure[]{
	\centering
	\includegraphics[height=3cm]{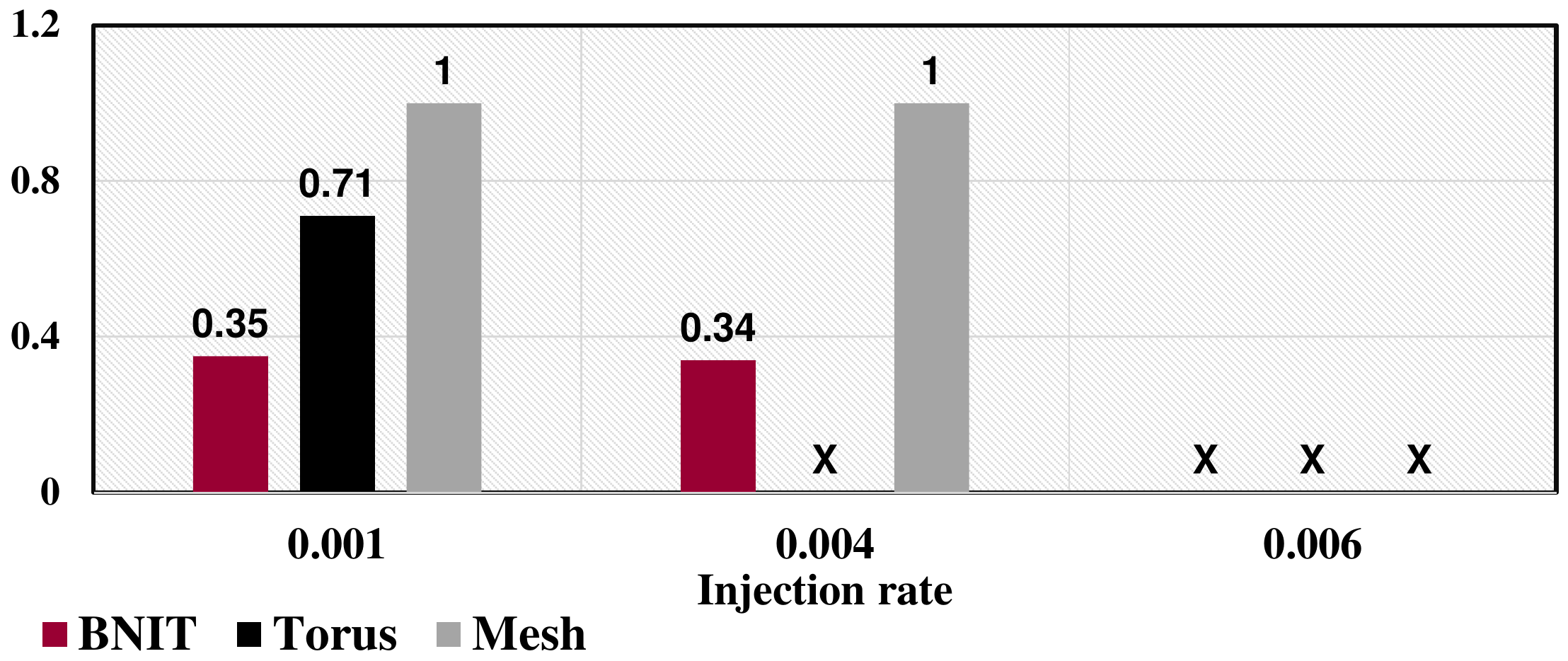}
	\label{fig:shuffle-lan}
}\\
\subfigure[]{
	\centering
	\includegraphics[height=3cm]{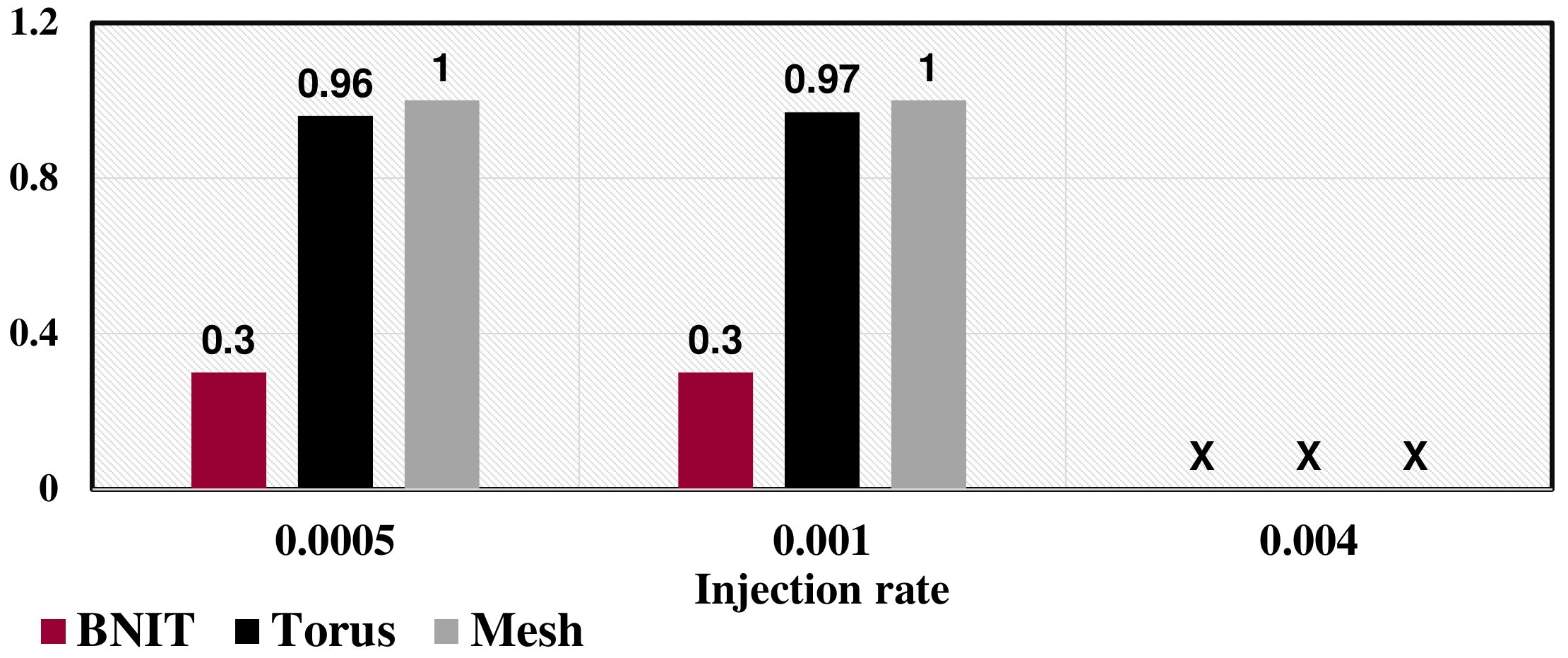}
	\label{fig:uniform-lan}
}
\hspace{5ex}
\subfigure[]{
	\centering
	\includegraphics[height=3cm]{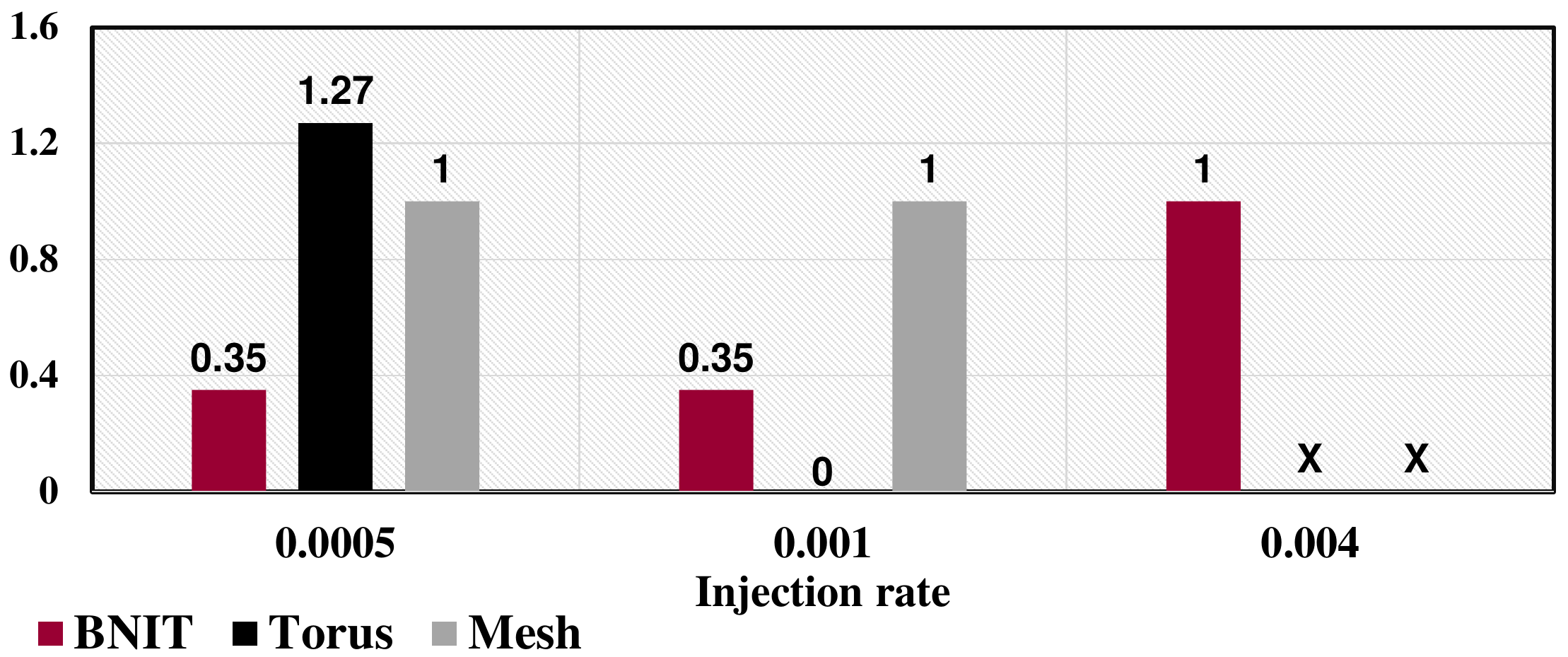}
	\label{fig:shuffle-lan}
}\\
\subfigure[]{
	\centering
	\includegraphics[height=3cm]{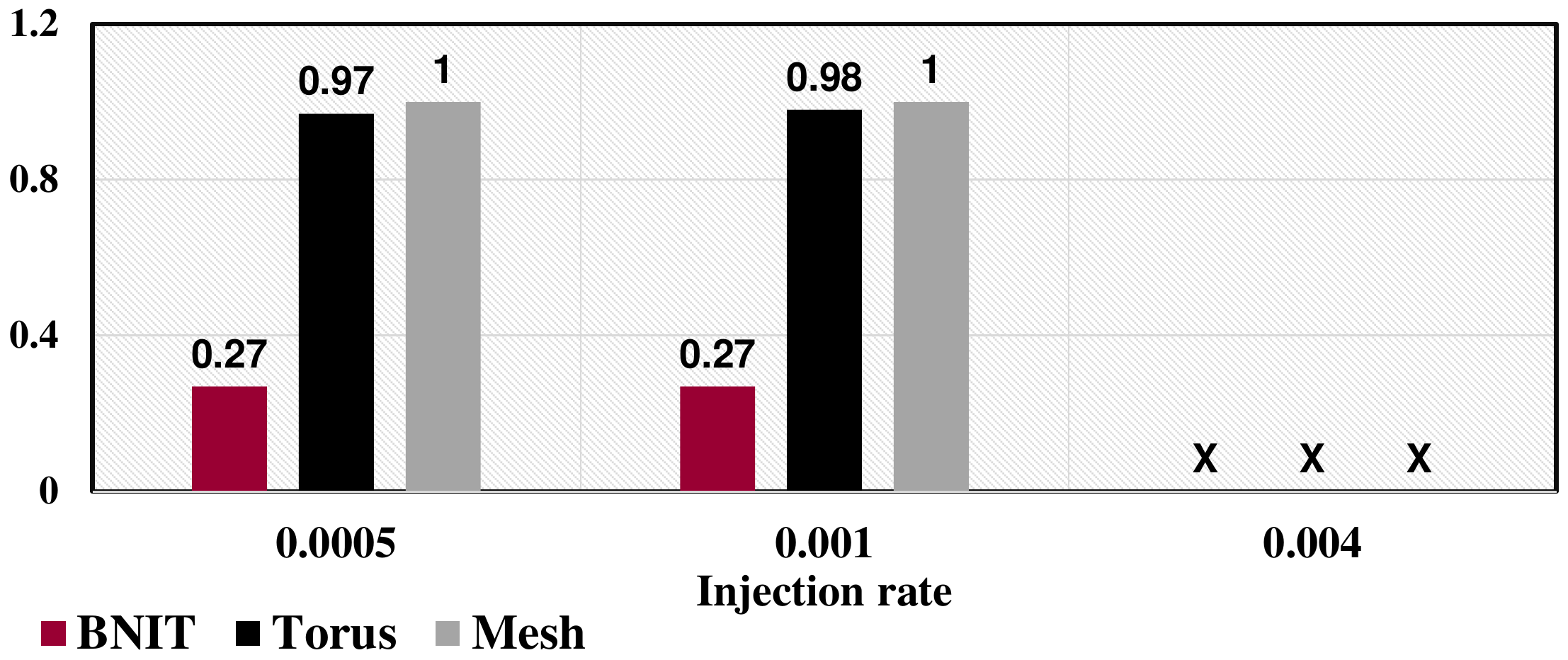}
	\label{fig:uniform-lan}
}
\hspace{5ex}
\subfigure[]{
	\centering
	\includegraphics[height=3cm]{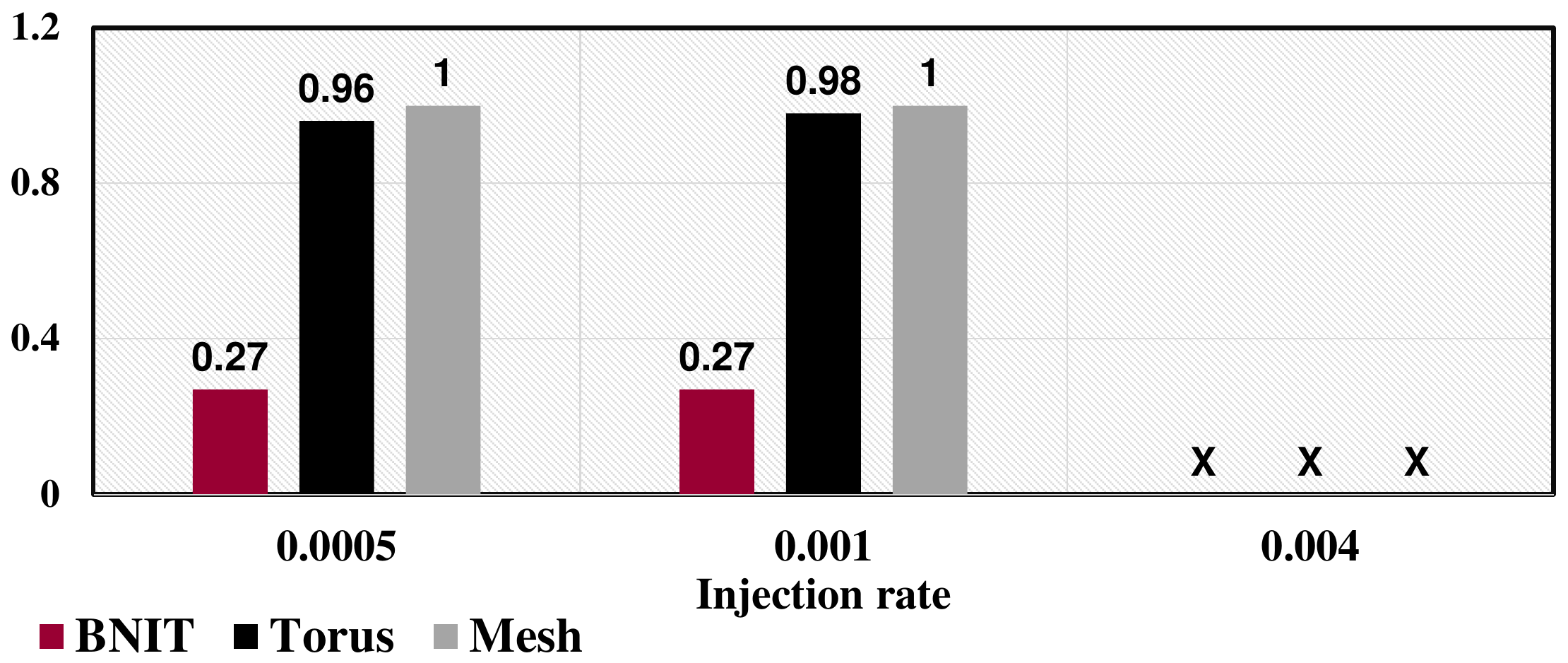}
	\label{fig:shuffle-lan}
}
	\caption{\revise{Normalized average communication latency for mesh, torus, and BNIT with 1024 cores at (a) uniform and (b) bitcomp traffic patterns, with 4096 cores at (c) uniform and (d) shuffle traffic patterns, and with 6400 cores at (e) uniform and (f) randperm traffic patterns.}}
\label{fig:booksim2}
\end{figure*}
Figure \ref{fig:booksim1} shows the power evaluation for different-scale mesh, torus, and BNITs at four different traffic patterns using $BookSim2$.
\majrev{Label ``$\times$" indicates that communication congestion occurs in the network simulation.}
Power consumption covers basic power and communication power consumption, which includes $wire\ power$, $read\& write\ power$ of buffers, $crossbar\ power$, and others.
With a low injection rate (0.0001 $\sim$ 0.001), \majrev{basic} power consumption dominates the communication architecture, so the total power of BNIT is about 20\% higher than others on account of some high-radix switches and long-range links.
With the increase of injection rate, the power consumption of mesh and torus is gradually larger and larger than that of BNIT, which mainly comes from more power consumption generated by \revise{frequent} $read\& write$ of buffers and traversal of crossbars.
When the injection rate \majrev{reaches 0.002 and above}, $read\& write\ power$ and $crossbar\ power$ of mesh, \majrev{which even account for about 25\% of total power consumption}, are \majrev{about 4} times that of BNIT.

Packet latency often has a direct impact on overall performance.
Figure \ref{fig:booksim2} depicts the normalized average communication latency of different-scale mesh, torus, and BNIT at four different traffic patterns.
In all cases, BNIT shows \majrev{an} average latency of at least \majrev{55}\% lower than mesh and torus.
For the deterministic packet routing, the lower average hop count of the routing paths makes sense and is reflected in network latency.

In summary, \majrev{under these traffic patterns with non-local communication, when the injection rate is slightly higher}, BNIT presents lower power consumption and average communication latency, which depends on the lower average hop count between any two nodes.
Hence, BNIT can be a promising solution for realizing global communication with strong coupling between cores.

\subsection{Performance Validation of Large-scale Applications Using $BookSim2$}
\label{sec:BOOKSIM}
Table \ref{tab:network-performance-booksim} and \ref{tab:network-performance-booksim2} depict the full evaluation of topologies with $1024\ (32*32)$, $4096\ (64*64)$, and $6400\ (80*80)$ cores, respectively.
We execute the greedy-based mapping algorithm \cite{Sawada2016TrueNorth} and WOAGA \cite{WangWOAGA} on mesh, torus, and BNIT for comparison.
Executing the proposed application mapping can directly provide average communication hop count \textbf{\#hop}, success \majrev{percentage \textbf{\#suc}}, and runtime \textbf{\#RT}.
Average latency \textbf{\#latency}, \majrev{basic} power consumption \textbf{\#PB}, \majrev{communication} power consumption \textbf{\#PC}, and total power consumption \textbf{\#PT} are derived from a cycle-accurate NoC simulation using $BookSim2$.
\majrev{The units of power consumption and average latency are respectively watts and cycles.}
\begin{table*}[htbp]
\centering
  \caption{Comparison of network performance with large-scale applications using $BookSim2$.}
  \label{tab:network-performance-booksim}
  \footnotesize
  \renewcommand{\arraystretch}{1.1}
  \renewcommand\tabcolsep{3pt}
  \begin{tabular}{|c|c|c|c|c|c|c|c|c|c|}
    \hline
    \textbf{Network size}&\textbf{Benchmark}&\textbf{Topo.+\ method} &\textbf{ \#hop } &\textbf{\#latency}&\textbf{ \#PB }&\textbf{ \#PC }&\textbf{ \#PT }&\textbf{ \#suc  }&\textbf{RT (s)}\\
    \hline
    \multirow{36}{*}{\textbf{$64*64 (4096)$}}&\multirow{14}{*}{\textbf{$G\_4096$}}
    &\multirow{2}{*}{\textbf{This work}}& \majrev{6.26}&\majrev{45.0}&\majrev{19.8}& \majrev{3.9}& \majrev{23.7}& \majrev{96.5}\%&\majrev{380}\\
    &&&-\majrev{61}\%	&-\majrev{43}\%	&+\majrev{26}\%	&-\majrev{11}\%	&+\majrev{17}\%	&+\majrev{86}\%	&-\majrev{2}\%\\
    \cline{3-10}
    &&\multirow{2}{*}{BNIT+greedy}& \majrev{6.65}&\majrev{49.0}&\majrev{19.8}&\majrev{4.5}& \majrev{24.3}& \majrev{96.0}\%&\majrev{424}\\
    &&&-\majrev{59}\%&-\majrev{38}\%&+\majrev{26}\%&	\majrev{+2}\%&	+\majrev{21}\%&+\majrev{85}\%&+\majrev{9}\%\\
    \cline{3-10}
    &&\multirow{2}{*}{Mesh+greedy}& \majrev{16.23}& \majrev{79.6}& \majrev{15.8}& \majrev{4.4}& \majrev{20.2}& \majrev{51.9}\%&\majrev{389}\\
        &&&1&1&1&1&1&1&1\\
    \cline{3-10}
    &&\multirow{2}{*}{Torus+greedy}& \majrev{13.5} & \majrev{119.8}& \majrev{16.2}&  \majrev{7.2} & \majrev{23.3}& \majrev{57.3}\%&\majrev{417}\\
        &&&-\majrev{17}\%&+\majrev{51}\%&+\majrev{2}\%	&+\majrev{63}\%&+\majrev{15}\%&+\majrev{10}\%&+\majrev{7}\%\\
    \cline{3-10}
    &&\multirow{2}{*}{BNIT+WOAGA}& \majrev{10.25}&\majrev{70.7} &\majrev{19.8}&\majrev{8.0}&\majrev{27.8}&\majrev{79.5}\% & \majrev{3251}\\
        &&&-\majrev{37}\%&-\majrev{11}\%&+\majrev{26}\%&+\majrev{81}\%&+\majrev{38}\%&+\majrev{53}\%&\majrev{$\times 74$}\\
    \cline{3-10}
    &&\multirow{2}{*}{Mesh+WOAGA}&\majrev{38.70}& - &\majrev{15.8}&\majrev{12.7}& \majrev{28.5}&\majrev{13.0}\%&\majrev{2888}\\
        &&&+\majrev{138}\%	&-&0\%	&+\majrev{189}\%	&+\majrev{41}\%&	-\majrev{75}\%	&\majrev{$\times 64$}\\
    \cline{3-10}
    &&\multirow{2}{*}{Torus+WOAGA}&\majrev{28.14}& - &\majrev{16.2}&\majrev{13.6}&\majrev{29.8}&\majrev{14.3}\%&\majrev{3096}\\
        &&&+\majrev{73}\%	&-&+\majrev{2}\%	&+\majrev{210}\%	&+\majrev{47}\%&-\majrev{72}\%&\majrev{$\times 70$}\\
    \cline{2-10}
            &\multicolumn{9}{|c|}{\centering{$\textbf{}$}}\\
    \cline{2-10}
      &\multirow{8}{*}{\textbf{$G\_3700$}}
      &\multirow{2}{*}{\textbf{This work}}& \majrev{6.00}& \majrev{43.8}& \majrev{19.8}&\majrev{4.8}& \majrev{24.6} & \majrev{97.0}\%&\majrev{342}\\
    &&&-\majrev{62}\%&-\majrev{44}\%&+\majrev{26}\%&	\majrev{-12}\%&	+\majrev{16}\%&+\majrev{87}\%&-\majrev{4}\%\\
    \cline{3-10}
    &&\multirow{2}{*}{BNIT+greedy}& \majrev{6.40}&\majrev{47.9}& \majrev{19.8}& \majrev{5.4}&\majrev{25.2}& \majrev{96.8}\%&\majrev{387}\\
    &&&-\majrev{59}\%&-\majrev{39}\%&+\majrev{26}\%&	\majrev{+1}\%&	+\majrev{19}\%&+\majrev{86}\%&+\majrev{9}\%\\
    \cline{3-10}
    &&\multirow{2}{*}{Mesh+greedy}&\majrev{15.79}&\majrev{78.6} &\majrev{15.8}&\majrev{5.5} &\majrev{21.3} & \majrev{52.0}\%&\majrev{356}\\
    &&&1&1&1&1&1&1&1\\
    \cline{3-10}
    &&\multirow{2}{*}{Torus+greedy}&\majrev{18.12} & \majrev{96.0} &\majrev{16.2}& \majrev{10.7} & \majrev{26.8}& \majrev{47.2}\%&\majrev{401}\\
    &&&+\majrev{15}\%&+\majrev{22}\%&+\majrev{2}\%&	\majrev{+96}\%&	+\majrev{26}\%&-\majrev{9}\%&+\majrev{13}\%\\
    \cline{2-10}
            &\multicolumn{9}{|c|}{\centering{$\textbf{}$}}\\
    \cline{2-10}
    &\multirow{8}{*}{\textbf{$VGG16$}}
    &\multirow{2}{*}{\textbf{This work}}&\majrev{5.62}&\majrev{30.5}&\majrev{19.8}&\majrev{2.2}&\majrev{22.0}&\majrev{99.3}\%&\majrev{353}\\
            &&&-\majrev{69}\%	&-\majrev{8}\%	&+\majrev{26}\%	&+\majrev{48}\%	&+\majrev{28}\%	&+\majrev{80}\%	&-\majrev{5}\%\\
    \cline{3-10}
    &&\multirow{2}{*}{BNIT+greedy}&\majrev{5.68}&\majrev{36.0}&\majrev{19.8}&\majrev{2.5}&\majrev{22.3}& \majrev{99.3}\%&\majrev{396}\\
           &&&-\majrev{68}\%	&+\majrev{8}\%	&+\majrev{26}\%	&+\majrev{68}\%	&+\majrev{29}\%	&+\majrev{80}\%	&+\majrev{7}\%\\
    \cline{3-10}
    &&\multirow{2}{*}{Mesh+greedy}&\majrev{17.92}&\majrev{33.3}& \majrev{15.8}& \majrev{1.5}& \majrev{17.3}& \majrev{55.2}\%&\majrev{370}\\
            &&&1&1&1&1&1&1&1\\
    \cline{3-10}
    &&\multirow{2}{*}{Torus+greedy}&\majrev{12.56} &\majrev{59.2}& \majrev{16.2}& \majrev{2.7}& \majrev{18.9}&\majrev{57.3}\%&\majrev{394}\\
           &&&-\majrev{30}\%	&+\majrev{78}\%&	+\majrev{2}\%	&+\majrev{82}\%&	+\majrev{9}\%&	+\majrev{3.8}\%&	+\majrev{6}\%\\
    \cline{2-10}
        &\multicolumn{9}{|c|}{\centering{$\textbf{}$}}\\
    \cline{2-10}
    &\multirow{8}{*}{\textbf{$D\_3980$}}
    &\multirow{2}{*}{\textbf{This work}}&\majrev{3.67}&\majrev{31.8}&\majrev{19.8}&\majrev{2.0}&\majrev{21.8}& \majrev{99.8}\%&\majrev{358}\\
            &&&-\majrev{7}\%&+\majrev{18}\%&+\majrev{26}\%&+\majrev{65}\%&+\majrev{29}\%&+\majrev{5}\%&-\majrev{4}\%\\
    \cline{3-10}
    &&\multirow{2}{*}{BNIT+greedy}& \majrev{3.68} &\majrev{32.3}&\majrev{19.8}&\majrev{2.2}&\majrev{22.0}&\majrev{99.8}\%&\majrev{401}\\
            &&&-\majrev{7}\%&+\majrev{20}\%&+\majrev{26}\%&+\majrev{82}\%&+\majrev{30}\%&+\majrev{5}\%&+\majrev{8}\%\\
    \cline{3-10}
    &&\multirow{2}{*}{Mesh+greedy}& \majrev{3.95}&\majrev{27.0}&\majrev{15.8}&\majrev{1.2}&\majrev{17.0}& \majrev{94.8}\%&\majrev{372}\\
        &&&1&1&1&1&1&1&1\\
    \cline{3-10}
    &&\multirow{2}{*}{Torus+greedy}& \majrev{5.70} & \majrev{59.2}& \majrev{16.2}&\majrev{2.6}&\majrev{18.8}& \majrev{99.5}\%&\majrev{410}\\
            &&&\majrev{+44}\%	&+\majrev{119}\%	&+\majrev{2}\%	&+\majrev{177}\%	&+\majrev{15}\%	&\majrev{-9}\%	&+\majrev{10}\%\\
    \hline
\end{tabular}
\end{table*}

In Table \ref{tab:network-performance-booksim} and \ref{tab:network-performance-booksim2}, success \majrev{percentage} means that the percentage of routing paths satisfying the constraints of hop count and link bandwidth.
BSOR \cite{BSOR} is used to allocate the routing path for each flow.
Compared with other ``topology+method", \majrev{in most cases}, the \majrev{generated brain-network-inspired} NoC design has a significantly lower average hop count, lower average latency, lower \majrev{communication} power consumption, and significantly higher success \majrev{percentage} for the different-scale benchmarks, even though our mapping method takes less runtime.
\majrev{As shown in Table \ref{tab:network-performance-booksim}, in addition to the application of $D\_3980$, on average, the brain-network-inspired NoC reduces the average hop count by 64\%, average latency by 32\%, and communication power consumption by 14\%, respectively, and increases success percentage by 84\%, compared with mesh-based NoC.
Besides, the total power consumption has increased by 20\% on average, due to the higher basic power consumption.}
\majrev{In particular, for} graph processing applications with a power-law and tightly coupled inter-core communication in Table \ref{tab:network-performance-booksim2}, the brain-network-inspired NoC architecture has up to \majrev{70\%} lower average hop count and \majrev{75\%} lower average latency than mesh-based NoC, \majrev{and the overall power consumption only increased by 10\%.}
\majrev{In addition, for $D\_3980$, which is composed of many small benchmarks, the traditional regular NoC may be more suitable, due to only local communication within each benchmark.}

\majrev{Although the brain-network-inspired topology has a very low average hop count, the mapping method will also obviously affect the overall communication performance of the communication architecture.}
On the whole, by comparing the three task mapping methods, \majrev{the proposed method is superior than others in all metrics and is suitable for BNIT.}
Specifically, compared to the greedy-based method \cite{Sawada2016TrueNorth} and WOAGA \cite{WangWOAGA}, our mapping method can obtain about \majrev{3}\% and \majrev{28}\% lower average hop count, and \majrev{10}\% and \majrev{35}\% lower \majrev{communication} power consumption, respectively.

\begin{figure}[htbp]
\small \centering
  \includegraphics[width=8.8cm]{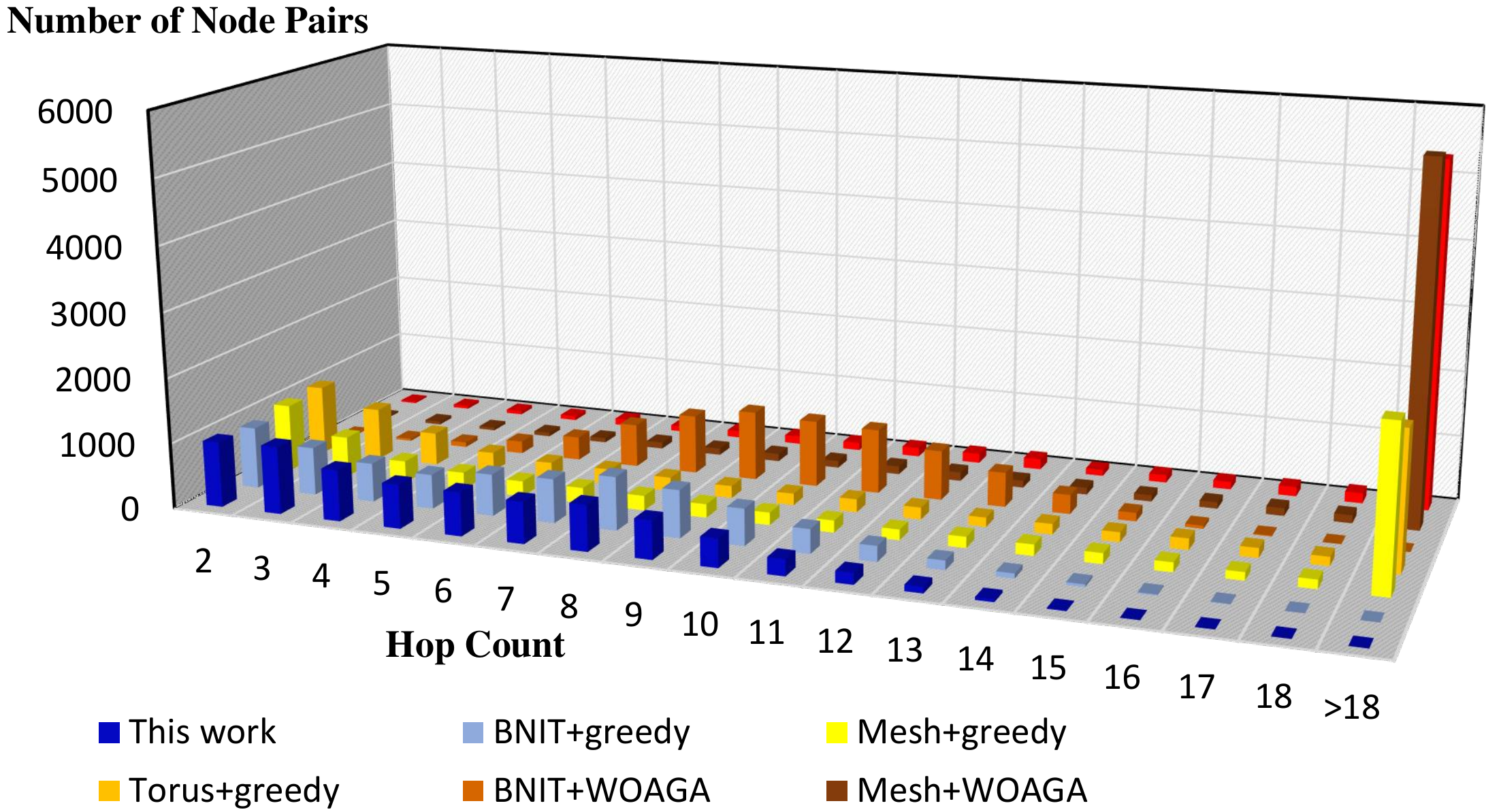}\\
  \caption{\revise{The hop count distribution of the routing path using different "topology+method" to process $G\_4096$ as shown in Table \ref{tab:network-performance-booksim}.}}
  \label{fig:hopdis}
\end{figure}

Figure \ref{fig:hopdis} depicts the hop count distribution of the routing paths using different topologies and mapping methods to process $G\_4096$ as shown in Table \ref{tab:network-performance-booksim}.
Mesh and torus have a very low success \majrev{percentage}, only around 50\%, and the hop count of most communication flows is very high and far beyond the constraint of hop count.
Thanks to the advantage of \majrev{the} extremely low average hop count of BNIT, almost all flows in NoC with any one of the three mapping methods have a small hop count for routing, \majrev{which also enables lower network communication latency.}

In summary, the generated brain-network-inspired NoCs have great advantages in three aspects, including the extremely low average hop count, low average latency, and high success \majrev{percentage}, for global communication.
When there are a large number of communication flows and strong inter-core coupling, \majrev{the lower communication power consumption of BNIT can compensate for its higher basic power consumption compared with mesh, and thus achieving a small increase in total power consumption.}
This also echoes the experiment results on BNIT under synthetic traffic patterns in Section \ref{sec:trafficpattern}.
The cycle-accurate simulations of applications demonstrate the effectiveness of this brain-network-inspired NoC for large-scale interconnections, especially as a promising domain-specific solution for graph processing \majrev{applications}.

\begin{table*}[htbp]
\centering
  \caption{\revise{Comparison of network performance with graph processing applications using $BookSim2$.}}
  \label{tab:network-performance-booksim2}
  \footnotesize
  \renewcommand{\arraystretch}{1.1}
  \renewcommand\tabcolsep{3pt}
  \begin{tabular}{|c|c|c|c|c|c|c|c|c|c|}
    \hline
    \textbf{Network size}&\textbf{Benchmark}&\textbf{Topo.+\ method} &\textbf{ \#hop } &\textbf{\#latency}&\textbf{ \#PB }&\textbf{ \#PC }&\textbf{ \#PT }&\textbf{\#succ}&\textbf{RT (s)}\\
    \hline
    \multirow{14}{*}{\textbf{$32*32(1024)$}}&\multirow{14}{*}{\textbf{$email$-$Eu$-$core$}}
    &\multirow{2}{*}{\textbf{This work}}&\majrev{6.40}&\majrev{50.3}&\majrev{4.7}&\majrev{2.5}&\majrev{7.2}&100\%&\majrev{4}\\
        &&&\majrev{-48}\%	&\majrev{-75}\%	&+\majrev{20}\%	&-\majrev{10}\%	&+\majrev{8}\%	&+\majrev{70}\%	&-\majrev{20}\%\\
    \cline{3-10}
    &&\multirow{2}{*}{BNIT+greedy}&\majrev{6.65}&\majrev{61.5}&\majrev{4.7}&\majrev{2.7}&\majrev{7.4}&100\%& \majrev{4}\\
        &&&-\majrev{46}\%	&\majrev{-70}\%	&+\majrev{20}\%	&\majrev{-6}\%	&\majrev{+9}\%	&+\majrev{70}\%	&-\majrev{20}\%\\
    \cline{3-10}
    &&\multirow{2}{*}{Mesh+greedy}&\majrev{12.36}&\majrev{207.4}&\majrev{3.9}&\majrev{2.8}&\majrev{6.7}&\majrev{58.8}\%& \majrev{5}\\
        &&&1&1&1&1&1&1&1\\
    \cline{3-10}
    &&\multirow{2}{*}{Torus+greedy}&\majrev{12.36}&\majrev{215.5}&\majrev{4.1}&\majrev{3.8}&\majrev{7.9}&\majrev{69.5}\%& \majrev{5}\\
        &&&\majrev{0}\%	&\majrev{+4}\%&	+\majrev{3}\%	&+\majrev{36}\%&	+\majrev{16}\%	&\majrev{+18}\%	&\majrev{0}\%\\
    \cline{3-10}
    &&\multirow{2}{*}{BNIT+WOAGA}&\majrev{7.78}&\majrev{65.8}&\majrev{4.7}&\majrev{3.3}&\majrev{8.0}&\majrev{99.2}\% & \majrev{383}\\
        &&&-\majrev{37}\%	&-\majrev{68}\%	&+\majrev{20}\%	&+\majrev{17}\%	&+\majrev{19}\%	&+\majrev{69}\%	&\majrev{$\times 76$}\\
    \cline{3-10}
    &&\multirow{2}{*}{Mesh+WOAGA}&\majrev{20.5}& - &\majrev{3.9}&\majrev{36.8}&\majrev{4.7}&\majrev{8.61}\%&\majrev{196}\\
        &&&+\majrev{66}\%	&- &\majrev{0}\%&	+\majrev{66}\%	&+\majrev{28}\%&	-\majrev{65}\%&\majrev{$\times 39$}\\
    \cline{3-10}
    &&\multirow{2}{*}{Torus+WOAGA}&\majrev{15.44}& - &\majrev{4.1}&\majrev{5.3}&\majrev{9.4}&\majrev{35.1}\%&\majrev{251}\\
            &&&+\majrev{25}\%	&-&\majrev{3}\%&	+\majrev{86}\%&	+\majrev{38}\%&	-\majrev{40}\%	&\majrev{$\times 50$}\\
    \hline
    \multicolumn{10}{|c|}{\centering{$\textbf{}$}}\\
    \hline
   \multirow{14}{*}{\textbf{$80*80(6400)$}}&\multirow{14}{*}{\textbf{$p2p$-$Gnutella08$}}
   &\multirow{2}{*}{\textbf{This work}}&\majrev{8.17}&\majrev{57.4}&\majrev{30.9}&\majrev{7.4}&\majrev{38.0}&\majrev{92}\%&\majrev{1351}\\
        &&&-\majrev{70}\%&	-\majrev{64}\%&+\majrev{26}\%&	-\majrev{23}\%&	+\majrev{12}\%&	+\majrev{237}\%&	-\majrev{2}\%\\
    \cline{3-10}
    &&\multirow{2}{*}{BNIT+greedy}&\majrev{8.28}&\majrev{59.7}&\majrev{30.9}&\majrev{7.8}&\majrev{38.7}&\majrev{92.3}\%& \majrev{1940} \\
        &&&-\majrev{69}\%&-\majrev{63}\%&+\majrev{26}\%&-\majrev{18}\%&	\majrev{13}\%&+\majrev{238}\%&	+\majrev{40}\%\\
    \cline{3-10}
    &&\multirow{2}{*}{Mesh+greedy}& \majrev{26.92}&\majrev{161.1}&\majrev{24.6}&\majrev{9.61}&\majrev{34.2}&\majrev{27.3}\%& \majrev{1382} \\
        &&&1&1&1&1&1&1&1\\
    \cline{3-10}
    &&\multirow{2}{*}{Torus+greedy}&\majrev{30.46}&\majrev{228.0}&\majrev{25.2}&\majrev{16.9}&\majrev{42.2}&\majrev{20.2}\%& \majrev{1865} \\
        &&&\majrev{+13}\%&+\majrev{42}\%&	+\majrev{3}\%&	+\majrev{76}\%&	+\majrev{23}\%&	-\majrev{26}\%&	+\majrev{35}\%\\
    \cline{3-10}
    &&\multirow{2}{*}{BNIT+WOAGA}& \majrev{11.00}&\majrev{75.6}&\majrev{30.9}&\majrev{11.0}&\majrev{41.9}&\majrev{69.5}\% & \majrev{12350}\\
        &&&-\majrev{59}\%&	-\majrev{53}\%&	+\majrev{26}\%&+\majrev{15}\%&	+\majrev{23}\%&+\majrev{155}\%&	\majrev{$\times 80$}\\
    \cline{3-10}
    &&\multirow{2}{*}{Mesh+WOAGA}&\majrev{50.91}& - &\majrev{24.6}&\majrev{19.9}&\majrev{44.5}&\majrev{6.7}\%&\majrev{10581}\\
        &&&+\majrev{89}\%&	-&\majrev{0}\%&+\majrev{107}\%&	+\majrev{30}\%&-\majrev{75}\%&	\majrev{$\times 67$}\\
    \cline{3-10}
    &&\multirow{2}{*}{Torus+WOAGA}&\majrev{37.82}& - &\majrev{25.2}&\majrev{21.8}&\majrev{47.1}&\majrev{7.3}\%&\majrev{11974}\\
            &&&+\majrev{41}\%&	- &+\majrev{3}\%&	+\majrev{127}\%&	+\majrev{38}\%&	-\majrev{73}\%&\majrev{$\times 77$}\\
    \hline
\end{tabular}
\end{table*}

\begin{figure}[htbp]
\small \centering
  \includegraphics[width=8cm]{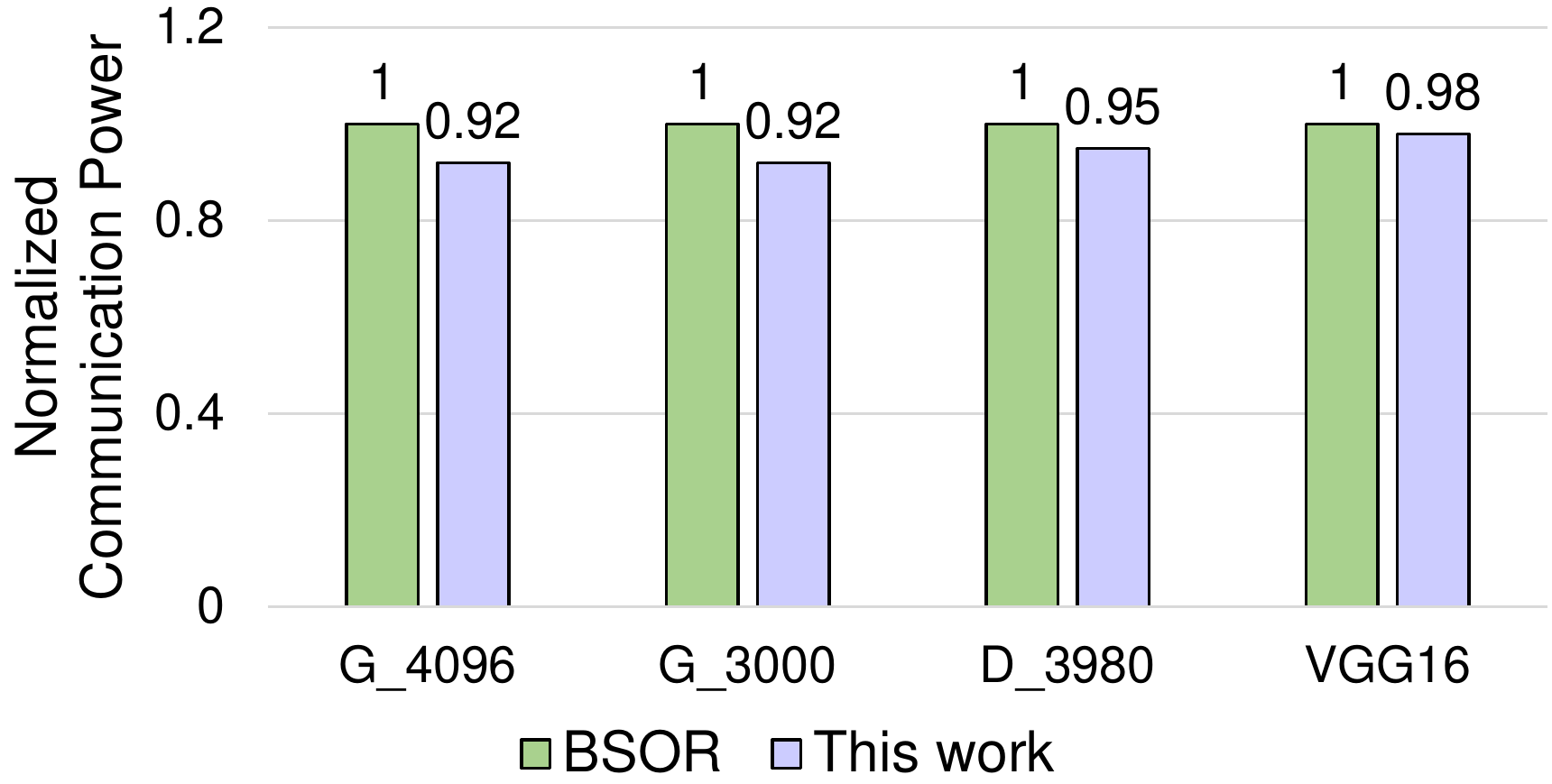}\\
  \caption{Normalized communication power of BSOR compared with the proposed routing algorithm for different benchmarks.}
  \label{fig:routing}
\end{figure}

In Figure \ref{fig:routing}, we can see that for the four benchmarks, compared with BSOR \cite{BSOR}, the \majrev{communication} power consumption based on the proposed Lagrangian relaxation method is reduced by about 7\% on average under a similar average hop count, and all routing paths meet the hop count and bandwidth constraint.
The results show the effectiveness of our routing algorithm.

\section{Conclusions}
\label{sec:conclusion}
\majrev{Regular interconnection topologies are simple and easy to implement, but they are not suitable for large-scale NoCs because of its extremely high average hop count and latency.}
In this paper, we propose to generate efficient brain-network-inspired interconnections for large-scale NoCs and address the large-scale application mapping problem for the brain-network-inspired NoC design.
\majrev{The simulation results show that, compared with conventional regular NoCs, the resulting brain-network-inspired NoC is a better solution to provide extremely low average hop count and low average latency for large-scale NoCs with global communication, especially in graph processing applications.}
In the future, we will further exploit to generate brain-network-inspired interconnections for large-scale NoCs on 3D ICs for higher integration and power efficiency.


\end{document}